\documentclass[conference,onecolumn]{IEEEtran}
\IEEEoverridecommandlockouts
\usepackage[switch]{lineno}
\usepackage{cite}
\usepackage{amsmath,amssymb,amsfonts}
\usepackage{algorithmic}
\usepackage{graphicx}
\usepackage{textcomp}
\usepackage{xcolor}
\usepackage[inline]{enumitem}
\usepackage{url}
\usepackage{booktabs}

\usepackage{setspace}
\begin{document}
% \linenumbers
\title{Repository Intelligence Graph: Deterministic Architectural Map for LLM Code Assistants\\
\thanks{This research was funded by the Blavatnik Family Foundation grant number 2953.}
}

\author{%
  Tsvi Cherny-Shahar\\
  Blavatnik School of Computer Science\\
  Tel Aviv University\\
  Israel\\
  \texttt{tsvic (at) mail.tau.ac.il}
  \and
  Amiram Yehudai\\
  Blavatnik School of Computer Science\\
  Tel Aviv University\\
  Israel\\
  \texttt{amiramy (at) tau.ac.il}
}

\maketitle

\begin{abstract}
Repository aware coding agents often struggle to recover build and test structure, especially in multilingual projects where cross language dependencies are encoded across heterogeneous build systems and tooling. We introduce the Repository Intelligence Graph (RIG), a deterministic, evidence backed architectural map that represents buildable components, aggregators, runners, tests, external packages, and package managers, connected by explicit dependency and coverage edges that trace back to concrete build and test definitions. We also present SPADE, a deterministic extractor that constructs RIG from build and test artifacts (currently with an automatic CMake plugin based on the CMake File API and CTest metadata), and exposes RIG as an LLM friendly JSON view that agents can treat as the authoritative description of repository structure.

We evaluate three commercial agents (Claude Code, Cursor, Codex) on eight repositories spanning low to high build oriented complexity, including the real world MetaFFI project. Each agent answers thirty structured questions per repository with and without RIG in context, and we measure accuracy, wall clock completion time, and efficiency (seconds per correct answer). Across repositories and agents, providing RIG improves mean accuracy by 12.2\% and reduces completion time by 53.9\%, yielding a mean 57.8\% reduction in seconds per correct answer. Gains are larger in multilingual repositories, which improve by 17.7\% in accuracy and 69.5\% in efficiency on average, compared to 6.6\% and 46.1\% in single language repositories. Qualitative analysis suggests that RIG shifts failures from structural misunderstandings toward reasoning mistakes over a correct structure, while rare regressions highlight that graph based reasoning quality remains a key factor.

\end{abstract}


%----------------------------------------------

\section{Introduction}
\label{sec:introduction}

Large language models are increasingly used as coding assistants and for repository level code completion in realistic multi-file projects \cite{jelodar_large_2025,shrivastava_repofusion_2023,cross_code_eval,liu_m2rc-eval_2024,zhang_repocoder_2023}. Modern development agents can open files, run tests, and invoke build tools, but they still struggle to form a reliable mental model of a project, especially when relevant information is scattered across many files and tools. In many real-world repositories, the build system, external dependencies, and test wiring are complex, and documentation is incomplete, out of date, or missing entirely. Agents are then forced to explore the repository through repeated tool calls and ad hoc pattern matching over files, which is slow and error-prone.

This problem is particularly evident in MetaFFI, a complex multi-lingual Foreign Function Interface framework that serves as the primary motivating example for this work \cite{cherny-shahar_metaffi-multilingual_2025}. MetaFFI combines multiple programming languages (for example Go, Python, and Java), uses a plugin-oriented architecture, and has deep dependency chains across languages \cite{cherny-shahar_metaffi-multilingual_2025}. In practice we observed that agents such as Claude Code~\cite{claude-code}, Cursor~\cite{cursor-cli}, and Codex~\cite{codex-cli} struggled to perform even relatively simple tasks in MetaFFI without repeated, manual explanations of the project structure. They often misidentified entry points, broke build configurations, or made incorrect assumptions about how components were wired together, while spending large numbers of tokens purely on exploration.

To address this gap, we propose the Software Program Architecture Discovery Engine (SPADE) and the Repository Intelligence Graph (RIG). RIG is a deterministic, build and test centered graph that represents a repository in terms of components, aggregators, runners, tests, external packages, and package managers. SPADE is an extractor that constructs RIG from build and test artifacts. It currently includes an automatic plugin for CMake that combines the CMake File API and CTest metadata and, when necessary, selectively parses CMake configuration (including commands invoked from custom targets). For other build systems, we currently use manually authored RIGs. RIG is serialized into a JSON view and provided in the context window of an LLM-based agent before tasks begin. The agent is instructed to treat RIG as the authoritative map of the build and test structure, so that it can answer structural questions by consulting RIG instead of reverse-engineering the build system. An overview of SPADE and RIG is given in Section~\ref{sec:novelty}. The RIG schema, entities, and extractor design are described in detail in Section~\ref{sec:spade-rig}. RIG and SPADE are available as open source at \url{https://github.com/Greenfuze/Spade}.

\textbf{Our goal in this work is not to compare alternative summary formats or to argue that JSON is inherently superior to other serializations such as Markdown or YAML.} JSON is a pragmatic choice: it is easy for SPADE to generate, validate, and post-process. The core contribution of RIG is \emph{not} the encoding format but the fact that it provides a deterministic, build and test derived architectural description that current agents simply do not have when they enter a repository. In the status quo, an agent must reconstruct components, dependencies, and test wiring through ad hoc file exploration and tool invocations. RIG instead supplies a build-system-grounded view that the agent can treat as authoritative. Any representation that preserved the same information would, in principle, be equally usable by an agent. SPADE and RIG contribute an automatic, reproducible way to obtain such an architecture-level view across heterogeneous repositories. Consequently, the meaningful baseline in our evaluation is the realistic setting in which agents operate \emph{without} any precomputed architectural knowledge, rather than hand-crafted or partially specified alternative summaries.

In order to quantify the impact of RIG on agent behavior, we construct a small corpus of eight repositories with varying complexity. The corpus includes simple synthetic projects such as a CMake-based \texttt{hello\_world}, medium-complexity multi-module and multi-service projects, and high-complexity repositories such as the MetaFFI codebase itself, an npm monorepo, and a Cargo-based compiler or interpreter. Repository complexity is measured using a build-oriented metric that combines component counts, language diversity, external packages, dependency depth, aggregators, and cross-language dependencies into a normalized score on a zero to one hundred scale, as described in Section~\ref{sec:complexity-metric}. This allows us to study how the benefit of RIG scales with repository complexity. The synthetic repositories are fully functional \footnote{All test repositories contain complete source code, build configurations, and dependency specifications. While designed to be buildable using standard tools (cmake, mvn package, cargo build, 
 npm run build), actual compilation was verified in development but not exhaustively tested across all build environments. Some implementation details are simplified for testing purposes (e.g.,
  UART drivers simulate hardware rather than interfacing with actual devices), which is standard practice for synthetic evaluation datasets.}. They are designed so that we know exactly what is in them and can control and explain their complexity.

We then design an evaluation in which three different commercial agents (Claude Code, Cursor, Codex) are asked to answer thirty structured questions per repository.
Questions are grouped into easy, medium, and hard difficulty levels, based on how likely agents are to answer them correctly without RIG and how much reasoning is required even when RIG is available. Each question has a precise expected answer format, and multiple acceptable answers are specified in advance. For each combination of repository and agent we run two configurations. In the baseline configuration the agent has access to the repository on disk and its usual tools, but no RIG. In the RIG configuration the agent receives the same environment plus the RIG JSON in its context. We measure both accuracy and wall-clock time from the moment we send the agent the prompt until it returns a complete answer. Because some agents do not report the number of tokens used, we treat completion time as a proxy for resource usage, and additionally report an efficiency metric defined as time per unit of score (seconds per score point).
Importantly, each agent is only compared to itself, with and without RIG, in order to isolate the contribution of RIG.
We do not compare agents against one another, as we are not interested in ranking models. The methodology and scoring procedures are described in Section~\ref{sec:methodology}, and the results are reported in Section~\ref{sec:evaluation-results}.

The contributions of this paper are as follows.

\begin{itemize}
  \item \textbf{Representation and extractor.} We introduce the Repository Intelligence Graph (RIG), a deterministic, build and test centered architectural graph over components, aggregators, runners, tests, external packages, and package managers, together with SPADE, an extractor that constructs RIG from build and test artifacts (Sections~\ref{sec:novelty} and~\ref{sec:spade-rig}). RIG is model and build-system-agnostic and evidence-backed. In our experiments it is serialized as JSON for convenience, but the contribution is the graph content rather than the specific encoding.

  \item \textbf{Benchmark and complexity metric.} We construct an evaluation corpus of eight repositories: seven synthetic but fully buildable projects plus the real MetaFFI repository. We define a normalized build-oriented complexity metric that combines the number of components, languages, external packages, dependency depth, aggregators, and cross-language edges (Section~\ref{sec:complexity-metric}). We release SPADE, the RIG schema, ground-truth RIGs, question sets, and scoring scripts to support reproducibility.

  \item \textbf{Empirical effects on agents.} Across eight repositories and three commercial agents, providing RIG in the context yields an average \emph{relative} accuracy improvement of \(12.2\%\) and an average reduction in completion time of \(53.9\%\), which corresponds to an average absolute reduction of 124.4 seconds per repository (mean of time without RIG minus time with RIG), compared to the same agents without RIG (Section~\ref{sec:evaluation-results}). Measured as efficiency (seconds per score point), the improvement is consistent across difficulty levels, ranging from \(62.3\%\) to \(66.4\%\) (Section~\ref{sec:evaluation-difficulty}).

  \item \textbf{Where RIG helps most.} We find that RIG is most beneficial on structurally complex settings. Medium and hard difficulty questions and high-complexity repositories see the largest improvements, with relative score gains up to \(28.6\%\) and efficiency gains up to \(79.2\%\) (Sections~\ref{sec:evaluation-complexity} and~\ref{sec:evaluation-difficulty}).

\end{itemize}

The rest of this paper is organized as follows.
Section~\ref{sec:novelty} introduces SPADE and the Repository Intelligence Graph at a conceptual level.
Section~\ref{sec:related-work} discusses related work on repository level graphs, retrieval and hierarchical methods, benchmarks, and static analysis with LLMs.
Section~\ref{sec:spade-rig} describes the RIG schema and the SPADE extractor in detail, including the CMake plugin and the JSON view used by agents.
Section~\ref{sec:methodology} presents the repository corpus, the complexity metric, the construction of RIGs and questions, and the experimental protocol.
Section~\ref{sec:evaluation-results} reports the evaluation results and analyzes how RIG affects accuracy, time, and efficiency across repositories, difficulty levels, and agents.
Section~\ref{sec:discussion} discusses the implications of these results for repository aware tools and benchmarks.
Section~\ref{sec:threats-to-validity} outlines threats to validity.
Section~\ref{sec:conclusion} concludes and outlines directions for future work.

\section{SPADE and RIG: Overview and Novelty}
\label{sec:novelty}

This section positions SPADE and the Repository Intelligence Graph (RIG) at a conceptual level and explains their role in our approach. At a high level, SPADE is a deterministic extractor that constructs RIG from build and test artifacts, and RIG is an evidence-backed architectural map of a repository that we inject into the context of LLM-based agents for downstream tasks. A detailed comparison with existing repository level graphs, retrieval methods, and benchmarks is given in Section~\ref{sec:related-work}. The concrete schema and extractor are described later in Section~\ref{sec:spade-rig}.

\subsection{Evidence-backed, build and test centered architectural map}

RIG treats a repository as a graph of evidence-backed build and test artifacts rather than as a bag of source files. Its nodes are components, aggregators, runners, test definitions, external packages, and package managers. Its edges record which components depend on which other components, which aggregators orchestrate which targets, which tests exercise which components, and which external packages are used. The schema and entity definitions are given in Section~\ref{sec:rig-schema}. In all cases, these nodes and edges are derived from concrete build and test evidence (for example, CMake File API outputs, test manifests, and build system metadata), or from their ground-truth equivalents in the manually authored RIGs.

This design emphasizes the build and test layer of a system as the backbone of its architecture. RIG does not attempt to represent abstract syntax trees, control flow graphs, or data flow graphs inside individual files. Instead it abstracts code into language-tagged components and focuses on how those components are built, linked, tested, and packaged, and how they depend on each other and on external packages. In this sense, RIG captures an application’s high-level architecture as seen through its build and test topology.

From the perspective of an LLM-based agent, RIG is an evidence-based architectural map of the repository. It answers questions such as which components exist, which components are built together, what the dependency structure looks like, which tests exist and what they cover, and which external libraries are used. It does so without asking the model to reverse-engineer the build system from scratch or to infer global structure solely from local file inspections.

As a concrete example, consider the \texttt{metaffi} repository and the question: ``What components does the MetaFFI top-level aggregator directly depend on?'' Without RIG, an agent must locate the relevant \texttt{CMakeLists.txt} files in the repository, identify the \texttt{MetaFFI} target, parse an \texttt{add\_custom\_target} declaration, interpret its \texttt{DEPENDS} clause, and distinguish aggregators from components while handling conditional logic, variable indirection and commented out code. With RIG, the same question is answered by reading the \texttt{depends\_on\_ids} field of the \texttt{MetaFFI} aggregator node in the JSON view and resolving those IDs to component names such as \texttt{metaffi-core}, \texttt{python311}, \texttt{openjdk}, and \texttt{go}. What is a multi-step build-file parsing task without RIG becomes a direct lookup over the graph.

\subsection{Deterministic extraction from build systems}

SPADE constructs RIG deterministically from build and test artifacts. For CMake projects, the SPADE CMake plugin uses the CMake File API and CTest metadata to discover targets, dependencies, tests, and external packages \cite{cmake,cmake_file_api_manual,ctest_json_manual}. Where the CMake File API does not expose a needed property, the extractor falls back to inspecting CMake files directly or, in rare cases, to looking at the underlying generated build system. The extraction pipeline is described in detail in Section~\ref{sec:spade-cmake}.

The key property is that no non-deterministic behavior is involved in building RIG in this work. Given a repository and configuration, SPADE runs CMake, reads the resulting machine-readable artifacts, and instantiates RIG entities using explicit rules. The same input yields the same RIG. Validation and fail-fast error handling enforce basic consistency, for example that all references resolve and that required metadata is present. When information cannot be recovered from available evidence, the extractor either omits the corresponding node or edge or marks the field as \texttt{UNKNOWN} rather than guessing.

The current CMake plugin is designed to handle standard CMake projects with executables, libraries, tests, and external packages, and it has been validated on the CMake-based repositories in our evaluation, including the multi-lingual MetaFFI project. It combines three sources of evidence: the CMake File API as the primary source for the build graph, direct inspection of \texttt{CMakeLists.txt} files for properties that are not exposed by the File API, and CTest JSON output for tests. We do not claim full coverage of all possible CMake features or patterns. More advanced capabilities such as comprehensive install location tracking, richer test framework detection, full resolution of complex generator expressions in advanced projects, and support for additional artifact types (for example Python wheels or npm packages) are currently architectural placeholders rather than fully implemented features.

This deterministic approach has a clear downside. Each build system requires explicit support and a dedicated extractor that implements many system-specific details and edge cases. For CMake we implement a plugin as described above. For other build systems (for example Maven, Meson, Cargo, and npm-based setups), we currently construct RIGs manually using the same schema. Projects that rely heavily on custom scripts or highly dynamic programmatic build descriptions, such as SCons-based builds, are particularly challenging, because much of the structure is encoded in general-purpose code rather than declarative configuration. These limitations motivate the future work on LLM-based pattern extraction using greedy decoding (or temperature zero - LLM0) discussed in Section~\ref{sec:conclusion}, which aims to reduce the need for per build system plugins without giving up determinism.

Deterministic extraction also simplifies the story for RIG updates. Because SPADE is a pure function of the repository state and configuration, regenerating a fresh RIG after a build system change is conceptually as simple as rerunning the extractor. In principle, agents could be given a tool that triggers SPADE, validates the resulting graph, and replaces the RIG in their context with an updated version when the repository evolves. We do not evaluate such an incremental update loop in this work, but the design of SPADE and RIG is compatible with this pattern.

We emphasize that the contribution of RIG is \emph{not} the choice of JSON as a representation format, but the automatic extraction of a deterministic, architecture-level view derived from build and test artifacts that current agents do not possess. Any equivalent serialization carrying the same information would, in principle, serve the same role.

\subsection{Precomputed, LLM-friendly context}

Once constructed, RIG is exposed to agents as a JSON document. The JSON view is a direct serialization of the RIG graph with a flat top-level structure and identifier-based cross-references, as described in Section~\ref{sec:rig-json-view}. The document contains arrays of components, aggregators, runners, tests, external packages, package managers, and evidence, along with repository and build system metadata.

The JSON view is designed so that an agent can execute typical repository understanding tasks by reading and combining RIG entries instead of spending tokens on directory walking, file content scanning, and ad hoc parsing of build files. In our evaluation, the full RIG JSON for a repository is injected into the agent context at the start of the session. The agent receives the questions and the RIG together and is instructed to treat the RIG as the authoritative description of the build and test structure. JSON is a pragmatic serialization choice: it is easy for SPADE to generate, validate, and post-process. Any equivalent serialization that preserved the same information could, in principle, serve the same role.

We experimented with more aggressive JSON compaction that replaced many repeated paths and names with very short alias tokens and a dense global mapping table. While this reduced byte and token size, it also made the RIG harder for the agent to interpret and led to lower scores in early experiments. In the configuration used for this paper we still apply a lighter form of compaction, where nodes are identified by stable IDs and relationships (for example, \texttt{depends\_on\_ids} or \texttt{external\_packages\_ids}) refer to those IDs rather than nesting full objects, which keeps the JSON relatively flat and avoids duplication. However, we keep field names and key structures descriptive, rather than aggressively aliasing them, to balance compactness with readability for the agent.

In practice, the RIG JSON view remains modest in size relative to modern LLM context windows. Across our eight repositories the average RIG size is 20{,}692 bytes, which corresponds to roughly 5{,}173 tokens using the common approximation of four bytes per token. Sizes range from 1{,}977 bytes for the simplest repository to 60{,}076 bytes for the largest RIG in the corpus. RIG size scales predictably with the build-oriented complexity metric in Section~\ref{sec:complexity-metric}: low complexity repositories produce RIGs on the order of a few kilobytes, medium complexity repositories around tens of kilobytes, and high complexity repositories up to about 60 kilobytes. Even the largest RIG in our corpus (about 15{,}000 tokens) fits comfortably within typical context budgets and, as shown in Section~\ref{sec:evaluation-results}, this overhead is associated with an average \emph{relative} score improvement of \(12.2\%\) and a \(53.9\%\) average reduction in completion time.

\subsection{Build-system-agnostic representation}

RIG is designed to be build-system-agnostic. The same schema can represent CMake-based projects, Meson-based projects, Maven multi-module applications, npm monorepos, Cargo-based compilers or interpreters, and Go projects.
 The entities described in Section~\ref{sec:rig-schema} are intentionally generic. Components are buildable artifacts independent of the build tool. Aggregators are orchestration targets. External packages and package managers model package ecosystems such as vcpkg, Maven, npm, Cargo, or pip.

In this work, the SPADE CMake plugin automatically constructs RIG for CMake repositories. For the other build systems in the evaluation corpus, RIGs are constructed manually using the same internal API that the CMake plugin uses. This ensures that all repositories, whether synthetic or real, share one canonical graph structure. Methodology details and the complexity metric used to characterize repositories appear in Section~\ref{sec:complexity-metric}.

This separation between schema and extractor is deliberate. It allows future work to add additional build-system plugins or alternative extractors while keeping the representation consumed by agents stable. It also makes it possible to compare different extraction methods, for example deterministic SPADE-based extraction and pattern-based extraction, by holding the schema constant.

A natural question is why we do not reuse an existing architecture description notation such as UML, possibly in a JSON-encoded form. RIG is deliberately not a general-purpose design language. UML is intended primarily for human-facing design and documentation, and its core abstractions (for example classes, components, deployment nodes) are not defined in terms of concrete build and test evidence. Mapping CMake targets, Meson build definitions, Maven modules, npm workspaces, or Cargo packages into UML diagrams would require many conventions that are specific to a given tool and project, and different mappings could encode different semantics for the same repository. The resulting models would be larger, more verbose, and less directly tied to what the build system actually does. In contrast, RIG is a compact, task-oriented schema whose nodes and edges are defined by how the build and test tools behave. It is intended as a machine-facing representation for agents rather than as a universal architecture notation. Nothing prevents generating UML-style diagrams from RIG for human consumption, but RIG itself is chosen as the canonical representation that SPADE produces and agents consume.

Concrete RIGs for our non-CMake repositories illustrate this Build-system-agnosticism. In the npm monorepo, compiled bundles such as \texttt{core.dist.js} or \texttt{auth-service.dist.js} are represented as \texttt{Component} nodes with \\\texttt{programming\_language = "typescript"} and \texttt{source\_files} pointing into the corresponding packages and services, while top-level scripts such as \texttt{@monorepo/build-all} and \texttt{@monorepo/test-all} appear as \texttt{Aggregator} nodes that depend on many components, and external npm dependencies such as \texttt{react} or \texttt{express} are \texttt{ExternalPackage} nodes linked to an npm \texttt{PackageManager}. In the Maven multi-module task manager repository, each module JAR and the final WAR file are \texttt{Component} nodes with \texttt{programming\_language = "java"}, the top-level reactor is an \texttt{Aggregator} node that depends on all module components, JUnit-based tests are \texttt{TestDefinition} nodes pointing to the components and test sources they exercise, and Maven dependencies appear as \texttt{ExternalPackage} nodes with their coordinates recorded via a \texttt{PackageManager}. The same schema applies to the Meson firmware, Cargo compiler, and Go microservices repositories, where components are firmware libraries, Rust crates, or Go services, aggregators are orchestration targets, tests are explicit \texttt{TestDefinition} nodes, and external packages and package managers describe the respective ecosystems.

\section{Related Work}
\label{sec:related-work}

Work on repository-scale coding with LLMs spans several strands that are relevant to this paper. These include repository level graphs and graph-aware models, repository-aware retrieval and training, hierarchical summarization and long-context structure, benchmarks for repository level reasoning, and static-analysis-assisted program understanding. We summarize the most closely related efforts and situate the Repository Intelligence Graph (RIG), introduced in Sections~\ref{sec:novelty} and~\ref{sec:spade-rig}, within this landscape.

\subsection{Repository level graphs and graph-aware models}

Several recent systems construct explicit graphs over code and use them to improve repository level reasoning. RepoGraph \cite{ouyang_repograph_2025} builds a fine-grained repository graph in which each node is a line of code and edges capture relationships such as definition and use, the graph is used as a plug-in module that retrieves ego subgraphs as context for repository level tasks such as SWE-Bench issue fixing and CrossCodeEval completion. CGM \cite{tao_code_2025} integrates a repository code graph directly into the model architecture by modifying attention to be graph-aware and mapping node attributes into the input space via a specialized adapter, and combines this with a graph retrieval pipeline. GraphCoder \cite{liu_graphcoder_2024} constructs a Code Context Graph that superimposes control flow, data dependence, and control dependence graphs over code statements and uses coarse-to-fine graph-based retrieval for repository level completion.

These approaches share two properties that are relevant here. First, they primarily operate at the code level, with graph nodes representing lines, statements, or functions rather than build or test artifacts. Second, they either couple the representation tightly to a specific model and training procedure, or focus on improving retrieval for completion tasks. In contrast, RIG is a build and test centered architectural graph that captures components, aggregators, runners, tests, and external packages (Section~\ref{sec:rig-schema}). It is external to the model and is designed as a reusable, model-agnostic map of build and test structure that can be injected into the context of any agent (Sections~\ref{sec:novelty} and~\ref{sec:rig-json-view}).

\subsection{Repository-aware retrieval and training}

Repository level retrieval and training methods exploit project structure to surface useful context or to train models that are more robust at repository scale. RepoCoder \cite{zhang_repocoder_2023} implements an iterative retrieval and generation loop that incorporates a similarity-based retriever and a pre-trained code language model to pull semantically similar code fragments across a project for repository level code completion. RepoFusion \cite{shrivastava_repofusion_2023} trains models with repository context for single-line code completion and shows that much smaller models can match or surpass larger ones when they are given carefully selected repository level context during training and inference.

These methods typically rely on similarity over code tokens and file-level metadata. They do not treat build and test artifacts as first-class, machine-readable structure. This can lead to over retrieval or to omission of critical context when similarity signals do not align with the true dependency graph. RIG is complementary, since it encodes build and test topology explicitly: components, aggregators, tests, external packages, and their dependency edges (Section~\ref{sec:rig-schema}). In principle, RIG can be used to restrict retrieval to dependency-closed slices, or to guide training data selection by mapping components to their sources, tests, and external packages, while keeping the LLM interface in the JSON form described in Section~\ref{sec:rig-json-view}.

\subsection{Hierarchical summarization and long context structure}

Beyond code-specific work, there is a line of research on hierarchical summarization and long-context processing that informs how architectural information can be organized. Dhulshette et al.\ \cite{dhulshette_hierarchical_2025} propose a two-step hierarchical approach for repository level code summarization tailored to business applications: smaller code units such as functions and variables are identified using syntax analysis and summarized with local LLMs, and those summaries are then aggregated into higher-level file and package summaries grounded in business context. Zhou et al. \cite{zhou_summarizing_2022} introduce a hierarchical code representation and hierarchical attention mechanism for source code summarization, operating at multiple code granularity levels to capture higher level structure beyond flat token or AST node sequences.

In natural language processing, HiStruct+ \cite{ruan_histruct_2022} explicitly formulates, extracts, encodes, and injects hierarchical document structure into an extractive summarization model and reports substantial ROUGE improvements on long scientific documents such as PubMed and arXiv articles. CHIME \cite{hsu_chime_2024} uses LLMs to produce hierarchical organizations of scientific studies for literature review, defining tree-structured hierarchies in which internal nodes are topical categories and leaves link to studies, and builds an expert-curated dataset of corrected hierarchies. NexusSum \cite{kim_nexussum_2025} introduces a multi-agent LLM framework for long-form narrative summarization that processes books, movies, and TV scripts through a structured, sequential pipeline with a hierarchical multi-LLM summarization stage. Classic hierarchical multi-document summarization by Christensen et al.\ \cite{christensen2014hierarchical} demonstrates that building and traversing hierarchical structures can scale summarization to larger document sets.

RIG is aligned with these ideas but focuses on build and test structure rather than textual or AST-based structure. It provides a ready-made hierarchy of components, aggregators, runners, and tests over which hierarchical agents or summarization pipelines can operate (Section~\ref{sec:rig-schema}). In this paper, RIG is used as a static map for structural question answering rather than as a basis for summarization, but the same representation could support hierarchical summarization or navigation of repositories in future work.

\subsection{Benchmarks and repository-scale evaluations}

Several benchmarks evaluate LLMs on repository level tasks. CrossCodeEval \cite{cross_code_eval} provides a diverse, multi-lingual cross-file code completion benchmark in which examples are constructed from real repositories and the benchmark pipeline retrieves and embeds relevant cross-file context into the model input to evaluate completion quality. M\textsuperscript{2}RC Eval \cite{liu_m2rc-eval_2024} scales this idea to a larger set of languages, covering 18 programming languages and providing fine-grained bucket-level and semantic-level annotations for repository level code completion scenarios. RepoExec \cite{hai_impacts_2025} focuses on repository level code generation: each task specifies natural language requirements and a set of essential code dependencies (contexts), and the benchmark evaluates whether the generated repositories compile and pass tests, including a Dependency Invocation Rate (DIR) metric that quantifies how well models use the provided cross-file contexts.

These benchmarks emphasize completion, generation, and executability, and they primarily target LLMs directly via curated textual prompts in which the relevant code context is explicitly selected and embedded by the benchmark pipeline. Build metadata is present in the underlying repositories they draw from, but it is not surfaced as an explicit graph that models can query or that evaluators analyze. Our evaluation is complementary. Instead of asking models to generate or complete code under a fixed prompt, we place agents in front of an actual filesystem checkout with their usual tools and add RIG as an additional, deterministic map of the build and test structure. We then ask those agents to answer structured architectural questions about components, dependencies, tests, and build order, and measure how deterministic build and test centered structure in the form of RIG changes their accuracy and latency (Sections~\ref{sec:methodology} and~\ref{sec:evaluation-results}).

\subsection{Static analysis and LLM-assisted program understanding}

A growing body of work explores how LLMs interact with static analysis and how they can assist with program understanding. Chapman et al.\ \cite{chapman_interleaving_2025} interleave static analysis with LLM prompting, feeding analyzer intermediate states to an LLM in order to infer missing error specifications and improve bug finding. The SEI study by Klieber and Flynn \cite{klieber_2024} reports that GPT-4 can triage static analysis alerts with explanations. Surveys and empirical studies such as Jelodar et al.\ \cite{jelodar_large_2025}, which reviews applications, models, and datasets for LLM-based source code analysis, and work on design pattern recognition with LLMs by Pandey et al.\ \cite{pandey_design_2025}, chart strengths and limitations of LLMs for code understanding and software analysis tasks.

These efforts typically operate at the level of code, control flow, and static analysis artifacts. Repository integration facts such as build artifacts, external packages, and test wiring are often implicit, and alerts are not always contextualized by which components are central in the build or transitively depended upon. RIG can complement such work by providing a view of the repository derived from build and test artifacts (Sections~\ref{sec:rig-schema} and~\ref{sec:spade-cmake}). For example, it can be used to prioritize alerts that affect components that are heavily depended on, or that are covered by critical tests.

\subsection{Summary}

In summary, prior work has convincingly shown that explicit structure, graphs, and repository level benchmarks can improve reasoning about large codebases. However, build and test artifacts are rarely modeled as a first-class, reusable graph that is external to the model and applicable across build systems. RIG, as constructed by SPADE (Section~\ref{sec:spade-rig}), fills this gap by providing a deterministic, build and test centered architectural map that is independent of any particular LLM or benchmark and that can be injected into an agent’s context in JSON form (Section~\ref{sec:rig-json-view}). Here, JSON serves as a convenient serialization rather than the core contribution, enabling agents to answer structural questions by consulting an evidence-backed map of build and test structure.

\section{SPADE and the Repository Intelligence Graph: Design and Structure}
\label{sec:spade-rig}

This section describes the design of the Repository Intelligence Graph (RIG) and the SPADE extractor that constructs RIG instances from build systems. RIG is the canonical representation that all experiments in this paper use, while SPADE is the concrete implementation that populates RIG for CMake projects. Methodology and evaluation procedures that use RIG are described in Section~\ref{sec:methodology}, and empirical results appear in Section~\ref{sec:evaluation-results}.

\subsection{Design goals and scope}
\label{sec:rig-design-goals}

RIG represents a repository as an evidence-backed graph over buildable components, orchestration targets, tests, and external packages, together with their dependencies.

\begin{itemize}
  \item \textbf{Build and test first.}
  RIG treats build and test artifacts as first-class nodes. It encodes what the build system actually compiles, links, or executes, which tests exist, and how they are wired to components and test frameworks.

  \item \textbf{Architectural rather than lexical.}
  RIG explicitly avoids lexical-token-level program structure. Nodes are components, aggregators, runners, test definitions, external packages, and package managers. There are no abstract syntax tree edges, control flow graphs, or data flow graphs in RIG. Those can be added by other tools if needed.

  \item \textbf{Model agnostic and build-system-agnostic.}
  RIG is external to any specific LLM or agent. The same schema can represent projects that use CMake, Meson, Maven, Cargo, npm, Go, or other build systems. SPADE currently implements an automatic extractor for CMake. For other build systems in this study we construct RIGs manually using the same schema.

  \item \textbf{Deterministic and reproducible.}
  For CMake repositories, SPADE derives RIG deterministically from build and test artifacts, with no LLM involvement. Given the same repository and configuration, the resulting RIG is uniquely determined by the extractor rules and by the build system outputs.

  \item \textbf{Evidence-backed.}
  Every graph fact is attached to evidence. Evidence items point back to concrete locations such as build files, configuration files, or test definitions. This makes it possible to trace RIG entries back to their origin.

  \item \textbf{LLM-friendly.}
  RIG is serialized into a JSON view that is suitable for inclusion in an LLM context window. Identifiers are short and stable, field names are regular, and cross-references use identifier sets. This keeps token budgets predictable while keeping the structure explicit. JSON is a pragmatic serialization choice, any equivalent format that preserves the same information could, in principle, serve the same role.
\end{itemize}

RIG is therefore an architectural map of the repository at the build and test level. SPADE is a deterministic extractor that populates this map from build system artifacts. In the evaluation (Sections~\ref{sec:methodology} and~\ref{sec:evaluation-results}), agents receive this map as a precomputed context and are scored on how well they use it. A complete formal specification of all entities and fields, including types and validation constraints, is given in the Appendix.

\subsection{Build and test centered architectural map}
\label{sec:rig-schema}

RIG treats a repository as an evidence-backed architectural map rather than as a bag of source files. Its core graph nodes are buildable components, orchestration targets, runners, and test definitions. External packages and package managers are represented as auxiliary entities linked from components. Its edges record which components depend on which other components, which aggregators orchestrate which targets, which tests exercise which components, and which external packages are used.

At the type level, the schema is implemented as a set of Pydantic models. The core graph nodes extend a common base class \texttt{RIGNode}. All such nodes share at least a string identifier \texttt{id}, a short human-readable \texttt{name}, a set \texttt{depends\_on\_ids} of identifiers of other nodes, and a set \texttt{evidence\_ids} of identifiers of evidence items. An intermediate base class \texttt{Artifact} further collects fields that are common to build artifacts, such as an output path relative to the repository root. Concrete entities extend these base classes with fields that match their role.

\paragraph{Repository and build system}

\texttt{RepositoryInfo} records repository level metadata such as the project name, repository root path, build directory, output directory, install directory, and the commands used to configure, build, test, and install the project. \texttt{BuildSystemInfo} records build system metadata such as the entrypoint build system name (for example CMake, Meson, Maven, Cargo, npm, or Go), a variant or generator name if applicable, and the selected build type. The top-level RIG object contains exactly one \texttt{RepositoryInfo} and one \texttt{BuildSystemInfo} instance, providing a canonical header for the graph.

\paragraph{Components and aggregators}

\texttt{Component} represents a code component that is built by the build system or interpreted at runtime. It extends \texttt{Artifact} and carries a \texttt{type} of enum type \texttt{ComponentType} (for example \texttt{executable}, \texttt{static\_library}, \texttt{shared\_library}, \texttt{package\_library} for Java class directories or Python wheels, \texttt{vm} for virtual-machine style artifacts such as JAR launchers, \texttt{interpreted}, or \texttt{unknown} when the extractor cannot determine the type). It also stores a lowercase \texttt{programming\_language} (such as \texttt{cxx}, \texttt{java}, \texttt{go}, \texttt{python}, \texttt{rust}, or \texttt{typescript}), a list of \texttt{source\_files} relative to the repository root, and references to any external packages it depends on. A component must either generate an artifact from source files, or represent source files directly if the language is interpreted.

\texttt{Aggregator} represents orchestration steps that do not generate new artifacts but group or invoke other build steps. Typical examples are targets like \emph{build all}, \emph{install all}, or \emph{test all}. Aggregators express their structure through \texttt{depends\_on\_ids}, which may reference any other RIG nodes, including components, other aggregators, runners, or tests, depending on how the underlying build system wires the workflow.

\paragraph{Runners and tests}

\texttt{Runner} represents entities that execute commands rather than generating artifacts. It extends \texttt{RIGNode} with an \texttt{arguments} list of command-line arguments and cross-references (\texttt{args\_nodes\_ids}) to RIG nodes that appear in those arguments. A test definition can point to a runner when tests are executed by commands such as \texttt{go test} rather than by a dedicated test executable component.

\texttt{TestDefinition} represents a test. It can point to a component or a runner that acts as the test executable, list components that implement the test harness, list components under test when this can be determined, record the source files that belong to the test, and name the test framework (for example \texttt{ctest}, \texttt{junit}, \texttt{pytest}, or \texttt{jest}). This allows RIG to model both compiled tests (common in C and C++) and command-based tests (common in dynamic or script-oriented ecosystems).

\paragraph{External packages and package managers}

\texttt{ExternalPackage} represents third-party dependencies such as vcpkg libraries, Maven artifacts, npm packages, Cargo crates, or Python packages. At minimum it has an \texttt{id} and a \texttt{name}, and in many cases it holds a reference to a \texttt{PackageManager}. \texttt{PackageManager} records how an external package is resolved, including its \texttt{name} (for example \texttt{vcpkg}, \texttt{maven}, \texttt{npm}, or \texttt{pip}) and the \texttt{package\_name} in that ecosystem. Components link to external packages through \texttt{external\_packages\_ids}, and external packages link to their package managers.

\paragraph{Evidence}

\texttt{Evidence} records the origin of graph facts. An evidence item has an \texttt{id} and at least one textual reference, such as a file name and line number pair (for example \texttt{packages/core/package.json:2}), or a call stack of build system locations. Every \texttt{RIGNode} must be associated with at least one evidence item. A node without evidence is considered invalid in a well-formed RIG. Nodes refer to evidence through their \texttt{evidence\_ids}. This not only supports debugging and manual inspection of the RIG, but also allows agents to jump directly from a node to the relevant location in the underlying build or configuration files.

\paragraph{RIG as a map}

From the perspective of an LLM-based agent, RIG is the map of the repository. It answers questions such as which components exist, which components are built together, what the dependency structure looks like, which tests exist, and which external libraries are used, without scanning source code or documentation and without asking the model to reverse-engineer the build system. For example, a hard question in our evaluation asks: ``What components does the MetaFFI top-level aggregator directly depend on?'' Without RIG, an agent must locate the corresponding custom target in the CMake files, understand that it is an orchestration target that does not produce an artifact, parse its \texttt{DEPENDS} clause, and resolve the referenced targets through variables and nested directories. With RIG, the same question reduces to looking up the MetaFFI aggregator node and reading its \texttt{depends\_on\_ids}, then resolving those identifiers to component names.

Table~\ref{tab:rig-entities} summarizes the core RIG entity kinds at this architectural level. The full schema, including all fields, types, and constraints (for example uniqueness of identifiers, allowed edge types, and validation error categories), is defined by the Pydantic models released with SPADE.

\begin{table}[t]
  \centering
  \caption{Core RIG entity kinds and their informal roles.}
  \label{tab:rig-entities}
  \begin{tabular}{ll}
    \toprule
    Entity type     & Informal role \\
    \midrule
    RepositoryInfo  & Repository level metadata and commands \\
    BuildSystemInfo & Build system identity and configuration \\
    Component       & Buildable artifact such as executable or library \\
    Aggregator      & Orchestration target that depends on other nodes \\
    Runner          & Command invocation that does not produce artifacts \\
    TestDefinition  & Test that links runners and components under test \\
    ExternalPackage & External dependency resolved by a package manager \\
    PackageManager  & Package ecosystem such as vcpkg, Maven, npm, Cargo, or pip \\
    Evidence        & Pointers back to build and configuration locations \\
    \bottomrule
  \end{tabular}
\end{table}

To keep RIG graphs predictable and machine-friendly, we impose a small set of normalization invariants on well-formed instances:

\begin{itemize}
  \item \textbf{Unique identifiers per collection.}
  Within a RIG, identifiers for components, aggregators, runners, tests, external packages, package managers, and evidence are unique inside their respective collections. All cross-references use the corresponding \texttt{*\_ids} fields rather than nested duplication.

  \item \textbf{Allowed node and edge types.}
  All graph nodes are instances of \texttt{Component}, \texttt{Aggregator}, \texttt{Runner}, or \\\texttt{TestDefinition}, each extending \texttt{RIGNode}. The \texttt{depends\_on\_ids} relation may point from any RIG node to any other RIG node, reflecting how real build systems can wire components, aggregators, runners, and tests together. Test-specific relations such as \texttt{test\_executable\_component\_id}, \texttt{test\_components\_ids}, and \\\texttt{components\_being\_tested\_ids} must reference existing components or runners of appropriate roles.

  \item \textbf{Evidence completeness.}
  Every \texttt{RIGNode} must be associated with at least one \texttt{Evidence} item, a node with an empty \texttt{evidence\_ids} set is considered invalid. Each \texttt{Evidence} entry must provide at least one concrete reference, either a file/line-style location or a build-system call stack. This guarantees that every fact in the graph can be traced back to specific build or configuration artifacts.

  \item \textbf{Validation and error reporting.}
  The extractor records violations (for example missing source files, broken dependencies, or circular dependencies) as \texttt{ValidationError} entries rather than silently repairing them. In the RIGs used for our experiments, SPADE is configured to fail-fast on severe validation errors and to avoid circular dependency cycles among build components, so that the resulting graphs can be interpreted as well-founded build and test structures.
\end{itemize}

\subsection{Internal representation and validation}
\label{sec:rig-internal}

The \texttt{RIG} class is the top-level container for a graph instance. It stores the repository information and build system information, and holds dictionaries that map identifiers to components, aggregators, runners, tests, external packages, package managers, and evidence.

Nodes are added through methods such as \texttt{add\_component}, \texttt{add\_aggregator}, \texttt{add\_runner}, and \texttt{add\_test}. Each of these methods registers the node in the corresponding dictionary and calls an internal helper that:

\begin{itemize}
  \item records evidence from the node and registers it in the global evidence collection,
  \item records any external packages attached to the node and updates \texttt{external\_packages\_ids},
  \item ensures that identifier sets such as \texttt{depends\_on\_ids} and other \texttt{*\_ids} fields are populated consistently.
\end{itemize}

In addition to identifier sets, components and tests can hold attached objects such as \texttt{external\_packages} or \\\texttt{test\_executable\_component}. Hydration passes reconstruct these object references from identifier sets when needed. For example, the component hydration pass iterates over all components and fills \texttt{external\_packages} lists from \\\texttt{external\_packages\_ids}. A similar pass reconnects tests with their executable components and with the components they exercise.

Before a RIG instance is used or persisted, SPADE calls \texttt{rig.validate()}. This method delegates to a dedicated \texttt{RIGValidator} that performs consistency checks and returns a list of \texttt{ValidationError} records. The SPADE code follows a fail-fast discipline: if validation returns any error of severity \texttt{ERROR}, the caller raises a \texttt{RIGValidationError} and does not write the graph out.

Validation checks currently include:

\begin{itemize}
  \item \textbf{Missing source files} (\texttt{missing\_source\_file}): every path in \texttt{Component.source\_files} must exist on disk, resolved relative to the repository root when applicable.
  \item \textbf{Broken dependencies} (\texttt{broken\_dependency}): all identifiers in \texttt{depends\_on\_ids} and other reference fields must point to existing nodes of the appropriate type.
  \item \textbf{Circular dependencies} (\texttt{circular\_dependency}): in configurations where the build system requires an acyclic build order, the dependency graph over RIG nodes must be free of cycles.
  \item \textbf{Duplicate node identifiers} (\texttt{duplicate\_node\_id}): identifiers for components, aggregators, runners, and tests must be unique within their collections.
  \item \textbf{Test relationships} (\texttt{missing\_test\_executable}, \texttt{test\_executable\_component\_not\_found}): each \\ \texttt{TestDefinition} that declares a test executable must reference an existing \texttt{Component} or \texttt{Runner}, and the corresponding node must be compatible with its role.
  \item \textbf{Evidence consistency} (\texttt{missing\_evidence}): every \texttt{RIGNode} must have at least one associated \texttt{Evidence} entry, and evidence must contain at least one concrete location (for example a file/line reference or a build-system call stack).
\end{itemize}

Each \texttt{ValidationError} carries a severity, category, message, and optional node, file path, line number, and suggestion fields, making validation failures actionable and traceable back to specific build or configuration artifacts.

The RIG implementation also supports persistence to SQLite through \texttt{save} and \texttt{load} methods. This allows graphs to be stored as durable artifacts and compared across runs using utilities that highlight differences between graphs in components, tests, and dependencies. In this paper, we use the SQLite form as an internal representation for inspection and regression checking rather than exposing it directly to agents.

\subsection{JSON view and LLM integration}
\label{sec:rig-json-view}

For use with LLM-based agents, SPADE exposes a JSON view of RIG. This view is generated by specialized \\\texttt{\_generate\_prompts\_*} methods that serialize each collection while excluding redundant object-level fields. The JSON view has the following properties.

\begin{itemize}
  \item \textbf{Flat top-level structure.}
  The root object contains keys \texttt{repo}, \texttt{build}, \texttt{components}, \texttt{aggregators}, \texttt{runners}, \texttt{tests}, \texttt{external\_packages}, \texttt{package\_managers}, and \texttt{evidence}. Relationships are represented by identifier sets such as \texttt{depends\_on\_ids} and \texttt{external\_packages\_ids}.

  \item \textbf{Short and regular identifiers.}
  Identifiers use short, regular prefixes such as \texttt{comp-1}, \texttt{agg-3}, \texttt{test-7}, \texttt{pkg-22}, \texttt{evidence-4}. This makes prompts easier to read and lowers token counts for references.

  \item \textbf{Predictable field names.}
  Field names are consistent across entities. Dependency sets are always called \texttt{depends\_on\_ids}. Evidence links are always \texttt{evidence\_ids}. External package links are always \texttt{external\_packages\_ids}. This regularity simplifies prompt templates.

  \item \textbf{No deep nesting.}
  Entities refer to each other by identifiers. Full objects are not nested inside other objects. This keeps the JSON size modest and makes it more likely that the entire RIG fits into the context window, while the identifier-based references avoid duplicating long paths or names.

  \item \textbf{Language tags.}
  Components carry a \texttt{programming\_language} field such as \texttt{cxx}, \texttt{java}, \texttt{go}, \texttt{python}, \texttt{rust}, or \texttt{typescript}. Agents can use this to reason about multi-lingual repositories and to restrict queries to specific languages.
\end{itemize}

We experimented with more aggressive JSON compaction where nearly all repeated strings (including paths and names) were mapped to very short aliases and a global mapping table was added to the JSON. In preliminary tests this reduced the raw byte and token size further, but also made the structure harder for agents to interpret and led to lower scores in our evaluation tasks. In the configuration used for this paper we therefore keep the lighter, identifier-based compaction described above: references go through short, stable IDs to keep the JSON flat and non redundant, but field names and most strings remain explicit.

RIG JSON sizes and approximate token counts for the repositories in our corpus, and their relationship to the normalized complexity scores, are quantified in Section~\ref{sec:rig-construction-questions} (Table~\ref{tab:rig-sizes-by-complexity}).

\subsection{CMake-based automatic extraction}
\label{sec:spade-cmake}

The current SPADE implementation provides an automatic extractor for CMake-based projects. The CMake plugin constructs a RIG instance from CMake and CTest artifacts, and, when required, from selected CMake input files.

At a high level, the extraction process for a CMake repository proceeds as follows.

\begin{enumerate}
  \item \textbf{Configure the project and invoke the CMake File API.}
  SPADE configures the project in an out-of-source build directory and uses the CMake File API \cite{cmake,cmake_file_api_manual} to obtain codemodel information, cache entries, toolchains, and file-level metadata.

  \item \textbf{Discover targets and artifacts.}
  The plugin enumerates CMake targets that produce artifacts such as executables and libraries. For each such target it creates a \texttt{Component} node, sets its component type and programming language, and records the relative path of the artifact.

  \item \textbf{Collect source files.}
  For each component, the plugin resolves the list of source files that contribute to the artifact. It follows the file lists that CMake has already computed via the File API, and, when necessary, traces CMake variables and helper functions that aggregate sources (for example \texttt{target\_sources} wrappers). Each source association is backed by evidence that points to the CMake definitions from which it was derived.

  \item \textbf{Capture dependencies.}
  The plugin extracts target-to-target dependencies from the File API and records them as \\ \texttt{depends\_on\_ids} edges between nodes. When dependencies involve third-party libraries or packages it creates \\ \texttt{ExternalPackage} nodes and links them to \texttt{PackageManager} entries such as vcpkg.

  \item \textbf{Record aggregators and custom targets.}
  CMake targets that do not produce artifacts but orchestrate other targets are mapped to \texttt{Aggregator} nodes. Their \texttt{depends\_on\_ids} sets point to the components, tests, runners, or other aggregators that they invoke.

  \item \textbf{Capture tests.}
  SPADE uses CTest metadata \cite{ctest_json_manual} to enumerate tests, identify their executables or command lines, and map them to \texttt{TestDefinition} nodes. Tests may point directly to compiled test components or to \texttt{Runner} nodes that represent commands such as \texttt{ctest} or language-specific test invocations.

  \item \textbf{Supplement with CMake parsing when needed.}
  In some cases the CMake File API and CTest data are not sufficient to reconstruct all relevant information. For example, complex custom commands or non-standard testing arrangements may only be apparent in the CMake configuration itself or in commands invoked from those targets. For such cases the plugin falls back to parsing selected fragments of the CMake configuration and, where applicable, records evidence that refers to external tools used by those custom commands. When information cannot be determined reliably, SPADE records explicit \texttt{UNKNOWN} values rather than guessing.

  \item \textbf{Attach evidence and validate.}
  Throughout extraction, the plugin records evidence entries that point back to CMake files, CTest configuration, or build files. Once the graph is populated it calls the RIG validator. If validation succeeds, the RIG is persisted and its JSON view is generated for downstream use.
\end{enumerate}

This extraction process is a research prototype. It targets the CMake features used by the repositories in this study and prioritizes correctness of the resulting RIG over full coverage of the CMake language. Extending SPADE to additional build systems and to more CMake features is part of future work.

For non-CMake repositories in the evaluation corpus, RIGs are constructed manually using the same internal API that the CMake plugin uses. Each manually authored RIG is passed through the same validator described in Section~\ref{sec:rig-internal}, and we cross-check its structure against the actual build and test behavior of the project (for example by comparing components and tests to build logs and by running the test suites). The MetaFFI ground-truth script is a concrete example that programmatically instantiates \texttt{RepositoryInfo}, \texttt{BuildSystemInfo}, \texttt{Component}, \texttt{TestDefinition}, \texttt{ExternalPackage}, and \texttt{PackageManager} objects and registers them with a \texttt{RIG} instance before validation (Section~\ref{sec:rig-construction-questions}).

\subsection{Current scope and limitations}
\label{sec:rig-limitations}

The current SPADE and RIG implementation focuses on the build and test layer. It does not contain language-specific code graphs (for example abstract syntax trees, control-flow graphs, or data-flow graphs). Instead, it abstracts code into components with language tags and records how components depend on each other and on external packages, and how tests exercise those components.

The automatic extractor currently supports CMake projects. The CMake plugin is engineered to cover the features needed by the CMake-based repositories in our evaluation: multi-lingual projects (for example C/C++, Java, Python, Go), custom CMake targets that act as orchestration aggregators, and several CTest patterns, including compiled unit-test components (for example C/C++ tests driven by CTest and doctest) and tests launched via runners that execute language-specific commands (for example Python or Java test runners). The plugin relies primarily on the CMake File API and CTest JSON metadata, and supplements them with targeted inspection of generated build files and, when necessary, selected CMake input fragments.

This coverage is sufficient for the selected CMake repositories, but it is not a complete implementation of the CMake language or ecosystem. Many advanced features, for example complex generator expressions, exotic generators, toolchain-specific extensions, and detailed install and packaging rules, are either outside the current scope or only partially supported. As a result, the evaluation results in Section~\ref{sec:evaluation-results} should be interpreted as evidence about the benefits of RIG given this subset of build systems and features, rather than as a complete characterization of all possible CMake projects.

For Meson, Maven, npm, Cargo, and Go-based repositories in this study, RIGs are constructed manually using the same internal API that the CMake plugin uses. These manually authored graphs are validated by the same \texttt{RIGValidator} described in Section~\ref{sec:rig-internal} and cross-checked against the actual build and test behavior of the projects, but they inevitably reflect the authors' understanding of each build configuration. We do not claim that this constitutes full coverage of each build system. Instead, it ensures that all repositories in the corpus share one canonical graph structure so that the effect of providing RIG to agents can be studied consistently.

A further limitation of the current approach is that each build system requires its own extractor, and highly customized build logic (for example bespoke scripts or build systems such as SCons) is difficult to support without substantial engineering effort. This motivates future work on constructing RIG using pattern matching over build artifacts with LLMs at temperature~0 (``LLM0''), which could reduce the need for hand-written plugins while preserving as much determinism as possible.

\section{Methodology}
\label{sec:methodology}

This section describes the repositories used in our evaluation, the complexity metric we use to characterize them, how we construct RIGs and questions, and how we run and score agent experiments. RIG itself and the SPADE extractor are described in Sections~\ref{sec:novelty} and~\ref{sec:spade-rig}. Here we treat them as given and focus on how they are used in the experimental setup.

\subsection{Repository corpus and complexity metric}
\label{sec:complexity-metric}

We evaluate SPADE and RIG on a corpus of eight repositories that cover a range of build systems, languages, and structural complexity. The corpus combines synthetic but fully functional projects and one real-world system (MetaFFI). Synthetic repositories are constructed so that their structure and contents are fully known in advance, which makes it possible to control and analyze complexity without relying on external documentation or reverse-engineering. Except for the real-world \texttt{metaffi} repository, these synthetic projects were implemented with the assistance of an LLM-based coding assistant and then documented, validated, and manually adjusted by the authors to match the intended architectures (see Section~\ref{sec:threats-to-validity}).

The corpus comprises:

\begin{itemize}
  \item \textbf{\texttt{hello\_world}} (low complexity). A simple CMake project with a single executable and a small utility library in C++. Single language, minimal dependencies, used as a baseline for RIG effectiveness.

  \item \textbf{\texttt{jni}} (medium complexity). A multi-lingual CMake project that integrates C++, Java, and Go via JNI. It has multiple build targets across three ecosystems and moderate cross language dependencies.

  \item \textbf{\texttt{metaffi}} (high complexity). The MetaFFI Foreign Function Interface framework, combining C++, Java, Go, and Python with a plugin architecture. It has deep dependency chains, multiple aggregators, extensive cross language integration, and modular runtime components.

  \item \textbf{\texttt{go}} microservices (medium complexity). A multi-lingual microservices-style repository centered on Go services with shared libraries and cross-language integration (C and JVM components). It includes eight buildable components, 11 external packages, and a maximum dependency depth of two.

  \item \textbf{\texttt{maven}} multi-module (medium complexity). A Maven multi-module Java application with ten interconnected modules, layered services (authentication, task management, notifications), persistence, and an API gateway. It exhibits a five-level dependency depth typical of enterprise Java systems.

  \item \textbf{\texttt{meson}} firmware (medium complexity). A Meson-based embedded firmware project that demonstrates code generation via \texttt{custom\_target}, build-time configuration via \texttt{configure\_file}, and subproject dependency management. It mixes C and C++ and has a two-level dependency depth.

  \item \textbf{\texttt{npm}} web application (high complexity). An npm monorepo with TypeScript, Rust (WASM), Go (native addons), and Python components across 22 buildable artifacts. It includes workspace management, cross-language FFI boundaries, a four-level dependency depth, and 22 external packages.

  \item \textbf{\texttt{cargo}} compiler or interpreter (high complexity). A Cargo based compiler or interpreter that exercises advanced Rust features including procedural macros, build scripts, C FFI integration, and a deep dependency hierarchy.
\end{itemize}

All eight repositories have RIG instances that conform to the schema in Section~\ref{sec:rig-schema}. For CMake-based projects \\(\texttt{hello\_world}, \texttt{jni}, and \texttt{metaffi}) RIG is produced automatically by SPADE (Section~\ref{sec:spade-cmake}). For the remaining repositories RIGs are constructed manually using the same internal API, ensuring a consistent representation across build systems.

Beyond covering different build systems and language ecosystems, the synthetic repositories are designed to emulate common real-world architectural patterns rather than toy examples. The Maven project follows a multi-module structure with layered services typical of enterprise Java systems. The Go project adopts a microservices layout with shared libraries and service-specific executables. The npm repository is a monorepo that combines TypeScript, Rust (WebAssembly), Go (native addons), and Python components under a shared workspace. The Cargo project models a compiler or interpreter with deep internal dependencies, procedural macros, and build scripts. This design ensures that agents face structurally realistic build and test topologies even when the repositories are synthetic.

Throughout the evaluation we always compare each agent to itself, with and without RIG, on the \emph{same} repository and question set. Our conclusions therefore depend on how RIG changes performance relative to each repository’s baseline, not on the absolute difficulty of any particular synthetic project. If an agent were to perform better or worse on different real-world repositories with similar build structures, the within-repository improvement (or lack of improvement) attributable to RIG would still be measured in the same way.

\textbf{Complexity metric.} To quantify how difficult a repository is to understand at the build and test level, we define a build-oriented complexity score. The score combines several dimensions that RIG exposes explicitly: number of components, number of programming languages, number of external packages, maximum dependency depth, number of aggregators, and presence of cross-language dependencies.

For each repository we compute a raw score
\[
\textit{raw\_score}
= 2 \cdot \textit{components}
+ 10 \cdot \textit{languages}
+ 3 \cdot \textit{packages}
+ 8 \cdot \textit{depth}
+ 5 \cdot \textit{aggregators}
+ \begin{cases}
  15 & \text{cross-language dependencies bonus}\\
  0  & \text{otherwise.}
\end{cases}
\]

The normalized complexity score is then
\[
\textit{normalized\_score} = \frac{\textit{raw\_score}}{\textit{max\_raw\_score}} \times 100,
\]
where \textit{max\_raw\_score} is the raw score of the most complex repository in the corpus. In our experiments this maximum is 221, obtained by the \texttt{cargo} compiler or interpreter repository.

We use thresholds on the normalized score to define three complexity bands:

\begin{itemize}
  \item \textbf{low complexity} for scores strictly below 30,
  \item \textbf{medium complexity} for scores between 30 and 70 (inclusive),
  \item \textbf{high complexity} for scores strictly above 70.
\end{itemize}

Under this metric \texttt{hello\_world} is low complexity, \texttt{jni}, \texttt{go}, \texttt{maven}, and \texttt{meson} are medium complexity, and \texttt{metaffi}, \texttt{npm}, and \texttt{cargo} are high complexity. Detailed calculations for each repository are provided in the supplementary material.

\subsection{RIG construction and question design}
\label{sec:rig-construction-questions}

For each repository we construct a RIG instance and a corresponding JSON view. The internal RIG is a strongly typed graph as described in Section~\ref{sec:rig-schema}. The JSON view is generated using the serialization described in Section~\ref{sec:rig-json-view} and is the form that agents consume.

For CMake-based repositories (\texttt{hello\_world}, \texttt{jni}, and \texttt{metaffi}), the RIG is created automatically by the SPADE CMake plugin (Section~\ref{sec:spade-cmake}). The plugin configures the project, reads CMake File API and CTest metadata, and instantiates \texttt{RepositoryInfo}, \texttt{BuildSystemInfo}, \texttt{Component}, \texttt{Aggregator}, \texttt{TestDefinition}, \texttt{ExternalPackage}, and \texttt{PackageManager} entities before validating the resulting graph.

For the Meson, Maven, npm, Cargo, and Go repositories we construct RIGs manually using the same Python API that the CMake plugin uses. For example, the Cargo ground-truth script shows how to construct a RIG programmatically by instantiating schema objects and registering them with a \texttt{RIG} instance. This manual construction ensures that the same schema and semantics apply across build systems even before SPADE plugins exist for all of them.

\textbf{Scope and role of the representation.} Our goal in this work is not to compare alternative summary formats or to argue that JSON is inherently superior to other serializations such as Markdown or YAML. The concrete choice of JSON is pragmatic: it is straightforward for SPADE to emit, easy to validate, and convenient for downstream tooling. The contribution of RIG is \emph{not} the syntax of the encoding, but the fact that it provides a deterministic architectural view derived from build and test artifacts that current LLM-based agents fundamentally lack. In the status quo, agents enter a repository with no explicit description of its build and test structure and must reconstruct components, dependencies, and test bindings through ad hoc file exploration and tool invocations. RIG supplies a build-system-grounded architectural description that agents can treat as authoritative. Any representation that preserved the same information would in principle be equally usable by an agent. What SPADE and RIG contribute is an automatic and reproducible way to construct such an architecture-level view across heterogeneous repositories. Accordingly, the meaningful baseline in our evaluation is the realistic setting in which agents operate \emph{without} any precomputed architectural knowledge, rather than hand-crafted or partially specified alternative summaries.

To quantify the overhead of including RIG in the agent’s context, Tables~\ref{tab:rig-sizes} and~\ref{tab:rig-sizes-by-complexity} report the size of the RIG JSON view per repository and aggregated by normalized complexity level. Across our eight repositories the average RIG size is 20{,}692 bytes, which corresponds to roughly 5{,}173 tokens using the common approximation of four bytes per token. Sizes range from 1{,}977 bytes for the simplest repository to 60{,}076 bytes for the largest RIG in the corpus. RIG size scales predictably with the build-oriented complexity metric introduced in Section~\ref{sec:complexity-metric}: low complexity repositories produce RIGs on the order of a few kilobytes, medium complexity repositories around tens of kilobytes, and high complexity repositories up to about 60 kilobytes. Even the largest RIG in our corpus (about 15{,}000 tokens) fits comfortably within typical context budgets, and, as shown in Section~\ref{sec:evaluation-results}, this overhead is associated with an average \emph{relative} score improvement of \(12.2\%\) and a \(53.9\%\) average reduction in completion time (124.4 seconds per repository).

\begin{table}[t]
  \centering
  \caption{RIG JSON size per repository. Sizes are measured in bytes and converted to approximate tokens using a four bytes per token approximation. Complexity is the normalized build-oriented score from Section~\ref{sec:complexity-metric}.}
  \label{tab:rig-sizes}
  \begin{tabular}{lcccc}
    \toprule
    Repository & Complexity & Level & RIG size (bytes) & RIG size (tokens) \\
    \midrule
    \texttt{hello\_world}       & 10.0  & LOW    & 1{,}977  & 494   \\
    \texttt{jni}         & 33.5  & MEDIUM & 5{,}967  & 1{,}491 \\
    \texttt{meson}  & 38.9  & MEDIUM & 7{,}816  & 1{,}954 \\
    \texttt{maven}                     & 47.5  & MEDIUM & 15{,}680 & 3{,}920 \\
    \texttt{go} (microservices)        & 54.3  & MEDIUM & 28{,}886 & 7{,}221 \\
    \texttt{metaffi}                   & 91.9  & HIGH   & 60{,}076 & 15{,}019 \\
    \texttt{npm}                       & 95.9  & HIGH   & 23{,}011 & 5{,}752 \\
    \texttt{cargo} (rholang)           & 100.0 & HIGH   & 22{,}121 & 5{,}530 \\
    \midrule
    Average                            &  --   &  --    & 20{,}692 & 5{,}173 \\
    \bottomrule
  \end{tabular}
\end{table}

\begin{table}[t]
  \centering
  \caption{Average RIG size by normalized complexity level.}
  \label{tab:rig-sizes-by-complexity}
  \begin{tabular}{lccc}
    \toprule
    Level & Avg size (bytes) & Avg tokens & Size range (bytes) \\
    \midrule
    LOW ($<$30)         & 1{,}977  & 494     & 1{,}977--1{,}977   \\
    MEDIUM (30--70)     & 14{,}587 & 3{,}647 & 5{,}967--28{,}886 \\
    HIGH ($>$70)        & 35{,}069 & 8{,}767 & 22{,}121--60{,}076 \\
    \bottomrule
  \end{tabular}
\end{table}

For each repository we then design a set of 30 evaluation questions. Questions are written so that they can be answered from repository structure and RIG content, and so that their expected answers can be specified precisely. The questions are not open-ended. They ask for specific numbers, component names, file paths, or dependency relations, and the expected answer format is part of each question.

The 30 questions are divided into three difficulty levels:

\begin{itemize}
  \item \textbf{Easy.} Questions that agents can answer reliably even without RIG, typically by simple file inspection or direct use of build commands. Examples include ``What is the project name?'' or ``How many \texttt{.cpp} files are in \texttt{src/}?'' RIG provides little benefit beyond convenience.

  \item \textbf{Medium.} Questions where CMake or other build syntax is nontrivial to parse and reason about, but where RIG contains a direct answer. Examples include ``What source files does \texttt{hello\_world} use?'' or ``Does \texttt{hello\_world} depend on \texttt{utils}?'' Without RIG agents must parse build files, with RIG they can perform a simple lookup. This is the regime where RIG is expected to have the largest impact.

  \item \textbf{Hard.} Questions that require reasoning even with RIG, such as computing build order or reverse dependencies. Examples include ``In what build order must components be built?'' or ``If \texttt{utils} fails to build, which components are affected?'' RIG makes the necessary information explicit but does not remove the need for reasoning over the graph.
\end{itemize}

The difficulty labels are assigned at design time based on the type and amount of reasoning required, rather than on any particular agent’s performance. Easy questions typically require only local file inspection or single-step lookups. Medium questions require understanding and following build or test structure that is explicit in RIG but implicit in the raw repository. Hard questions require multi-step reasoning over the dependency graph, such as computing build order or reverse dependencies. In Section~\ref{sec:evaluation-results} we report how accuracy varies across these difficulty bands with and without RIG.

Each question is also tagged with one or more semantic categories that reflect the type of reasoning required, such as \emph{build system structure} (targets, generators, build commands), \emph{dependency analysis} (incoming and outgoing edges between components and aggregators), \emph{external packages} (package managers and third-party libraries), and \emph{testing} (test definitions, test frameworks, coverage of components). These tags are used when designing the question sets to ensure coverage across build, dependency, and testing concerns, and they help interpret the aggregated results in Section~\ref{sec:evaluation-results}, even though we primarily report metrics averaged over difficulty and repositories.

We validate questions and expected answers manually by running pilot trials and by examining agent outputs. If agents produce correct information in an unexpected but reasonable format, the question or answer specification is refined to avoid penalizing such cases.

\subsection{Scoring and performance metrics}
\label{sec:scoring-metrics}

For each run, we score agent answers and compute performance metrics. Scoring is binary at the question level. A question is counted as correct if the agent's answer matches one of the acceptable answers for that question, allowing for minor formatting differences such as extra whitespace. Otherwise it is counted as incorrect. We do not distinguish between hallucinations and other forms of error. Any incorrect answer, whether it invents a plausible but wrong fact or omits required information, is simply marked wrong.

Each run, therefore, yields a raw score between 0 and 30 (the number of correctly answered questions) and a corresponding accuracy (raw score divided by 30). We report:

\begin{itemize}
  \item \textbf{Raw score}, the number of correctly answered questions (0-30).
  \item \textbf{Accuracy}, the fraction of correctly answered questions.
  \item \textbf{Time}, the end-to-end completion time in seconds, measured from the moment the agent receives the full set of questions until it returns answers for all of them in a single run. We exclude any external startup overhead (for example, model warm-up or session initialization) from this measurement.
  \item \textbf{Efficiency}, defined as seconds per score point (time divided by raw score). Lower values indicate better efficiency.
\end{itemize}

When comparing runs with and without RIG for the same agent and repository we derive:

\begin{itemize}
  \item \textbf{Score improvement (relative)}, the percentage increase in accuracy with RIG relative to accuracy without RIG.
  \item \textbf{Time reduction}, the percentage decrease in completion time when RIG is used.
  \item \textbf{Efficiency improvement}, the percentage decrease in the seconds-per-score-point metric when RIG is used (that is, how much less time is spent per correct answer).
\end{itemize}

When reporting averages across repositories, we compute relative improvements per repository and then take a simple mean across repositories. We treat these as within-agent effects. That is, we compare each agent to itself with and without RIG rather than ranking agents against one another. This isolates the impact of RIG as an additional source of structured context.

To improve scoring robustness, we designed the question set and acceptable answer lists iteratively (Section~\ref{sec:rig-construction-questions}). During pilot runs we inspected agent outputs and refined answer specifications whenever agents produced correct information in an unexpected but reasonable format (for example, alternative phrasings, synonymous component names, or semantically equivalent values such as \texttt{yes}/\texttt{y}/\texttt{true}). For the final evaluation, we spot-checked a sample of scored interactions across repositories and agents to confirm that the automatic string-based scoring aligned with human judgment in borderline cases. When discrepancies were found, we updated the answer specification and re-scored all runs for the affected question using the same deterministic scorer.

Aggregated results by repository, complexity band, question difficulty, question category, and agent are reported and analyzed in Section~\ref{sec:evaluation-results}.

\section{Evaluation Results}
\label{sec:evaluation-results}

This section reports the quantitative impact of RIG on agent performance as defined in Section~\ref{sec:scoring-metrics}. We first present repository level accuracy, time, and efficiency with and without RIG, then examine how these metrics vary with repository complexity, question difficulty, semantic category, and agent implementation. Higher scores indicate better accuracy, shorter wall-clock time indicates faster completion, and efficiency is measured as seconds per score point (lower values are better).

\subsection{Repository level accuracy}
\label{sec:evaluation-repo-accuracy}

Table~\ref{tab:repo-scores} summarizes accuracy with and without RIG for each repository, averaged over all three commercial agents. The complexity column reports the normalized build-oriented complexity score and band (low, medium, high) from Section~\ref{sec:complexity-metric}. Repositories are ordered by increasing complexity.

\begin{table}[t]
  \centering
  \caption{Repository level accuracy with and without RIG, averaged over all agents. Complexity in parentheses is the normalized score from Section~\ref{sec:complexity-metric}. Repositories are ordered by increasing complexity. \emph{Improvement} is the relative change in accuracy computed from the underlying (unrounded) accuracies.}
  \label{tab:repo-scores}
  \begin{tabular}{lcccc}
    \toprule
    Repository & Complexity & Accuracy w/o RIG & Accuracy with RIG & Improvement \% \\
    \midrule
    \texttt{hello\_world} & LOW (10.0)   & 98.9\% & 100.0\% & +1.1\%  \\
    \texttt{jni}          & MED (33.5)   & 91.1\% & 98.9\%  & +8.6\%  \\
    \texttt{meson}        & MED (38.9)   & 88.9\% & 97.8\%  & +10.2\% \\
    \texttt{maven}        & MED (47.5)   & 95.6\% & 98.9\%  & +3.7\%  \\
    \texttt{go}           & MED (54.3)   & 80.0\% & 96.7\%  & +21.5\% \\
    \texttt{metaffi}      & HIGH (91.9)  & 82.2\% & 95.6\%  & +16.3\% \\
    \texttt{npm}          & HIGH (95.9)  & 67.8\% & 84.4\%  & +24.5\% \\
    \texttt{cargo}        & HIGH (100.0) & 90.0\% & 100.0\% & +11.4\% \\
    \bottomrule
  \end{tabular}
\end{table}

% Optional: visual version of the same data.
% \begin{figure}[t]
%   \centering
%   \includegraphics[width=\linewidth]{analysis_images/repository_comparison_scores.png}
%   \caption{Repository level accuracy with and without RIG (all agents combined).}
%   \label{fig:repo-scores-fig}
% \end{figure}

Across all eight repositories, the unweighted mean \emph{relative} accuracy improvement (averaging the last column of Table~\ref{tab:repo-scores}) is \(12.2\%\). Improvements range from \(+1.1\%\) on the simple \texttt{hello\_world} project, where all agents are already near ceiling without RIG, to \(+24.5\%\) on the high-complexity \texttt{npm} monorepo. All repositories see non-negative accuracy changes when RIG is available. There are no cases where accuracy decreases.

\subsection{Repository level time and efficiency}
\label{sec:evaluation-repo-time}

Table~\ref{tab:repo-time} reports wall-clock completion times with and without RIG, averaged over agents, together with absolute and relative time savings per repository. Time is measured from the moment the full question set is submitted to the agent until it produces its final answers, including all intermediate tool calls, but excluding any one-off startup or warm-up activity outside the run. Repositories are ordered by increasing complexity.

\begin{table}[t]
  \centering
  \caption{Repository level completion times with and without RIG (all agents combined). Time is wall-clock seconds from prompt to completion. Repositories are ordered by increasing complexity.}
  \label{tab:repo-time}
  \begin{tabular}{lcccc}
    \toprule
    Repository & Time w/o RIG [s] & Time with RIG [s] & Time Saved [s] & Reduction \\
    \midrule
    \texttt{hello\_world} & 56.0  & 44.3  & 11.7  & 20.8\% \\
    \texttt{jni}          & 122.7 & 47.0  & 75.7  & 61.7\% \\
    \texttt{meson}        & 170.3 & 73.3  & 97.0  & 56.9\% \\
    \texttt{maven}        & 104.7 & 67.7  & 37.0  & 35.4\% \\
    \texttt{go}           & 230.0 & 82.3  & 147.7 & 64.2\% \\
    \texttt{metaffi}      & 258.0 & 109.0 & 149.0 & 57.8\% \\
    \texttt{npm}          & 440.0 & 114.0 & 326.0 & 74.1\% \\
    \texttt{cargo}        & 251.3 & 100.0 & 151.3 & 60.2\% \\
    \bottomrule
  \end{tabular}
\end{table}

Across repositories, the mean relative time reduction is \(53.9\%\), corresponding to an average savings of \(124.4\) seconds per repository: agents typically complete the same 30 question workload in a little under half the time when RIG is available. Time savings are largest on the more complex repositories, with \texttt{metaffi}, \texttt{npm}, and \texttt{cargo} all seeing reductions above \(57\%\).

% same as table
% \begin{figure}[t]
%   \centering
%   \includegraphics[width=\linewidth]{analysis_images/time_savings_cascade.png}
%   \caption{Time savings per repository with RIG. Bars show time without RIG, time with RIG, and absolute savings.}
%   \label{fig:time-savings-cascade}
% \end{figure}

Table~\ref{tab:repo-efficiency} summarizes the combined effect on efficiency, measured as seconds per correct answer. The efficiency improvement column is computed relative to the baseline without RIG, and the \emph{is multi-lingual} column indicates whether the agent must reason about multiple distinct language ecosystems in that repository.

\begin{table}[t]
  \centering
  \caption{Repository level score and efficiency improvements with RIG. Efficiency is defined as seconds per correct answer, positive percentages indicate lower time per correct answer. The \emph{is multi-lingual} column reflects whether the repository contains multiple distinct language ecosystems from the agent's perspective. Repositories are ordered by increasing complexity.}
  \label{tab:repo-efficiency}
  \begin{tabular}{lccccc}
    \toprule
    Repository & Complexity & Is multi-lingual & Score Improvement & Efficiency Improvement \\
    \midrule
    \texttt{hello\_world} & LOW (10.0)  & false          & +1.1\%  & +21.7\% \\
    \texttt{jni}          & MED (33.5)  & true           & +8.6\%  & +64.7\% \\
    \texttt{meson}        & MED (38.9)  & false          & +10.2\% & +60.9\% \\
    \texttt{maven}        & MED (47.5)  & false          & +3.7\%  & +37.5\% \\
    \texttt{go}           & MED (54.3)  & true           & +21.5\% & +70.4\% \\
    \texttt{metaffi}      & HIGH (91.9) & true           & +16.3\% & +63.6\% \\
    \texttt{npm}          & HIGH (95.9) & true           & +24.5\% & +79.2\% \\
    \texttt{cargo}        & HIGH (100.0)& false          & +11.4\% & +64.2\% \\
    \bottomrule
  \end{tabular}
\end{table}

Averaging the efficiency improvements in Table~\ref{tab:repo-efficiency} over all eight repositories yields an average efficiency gain of \(57.8\%\), measured as a reduction in seconds per correct answer. The high-complexity repositories \texttt{metaffi}, \texttt{npm}, and \texttt{cargo} see efficiency improvements of \(63.6\%\), \(79.2\%\), and \(64.2\%\), respectively.

\begin{figure}[t]
  \centering
  \includegraphics[width=\linewidth]{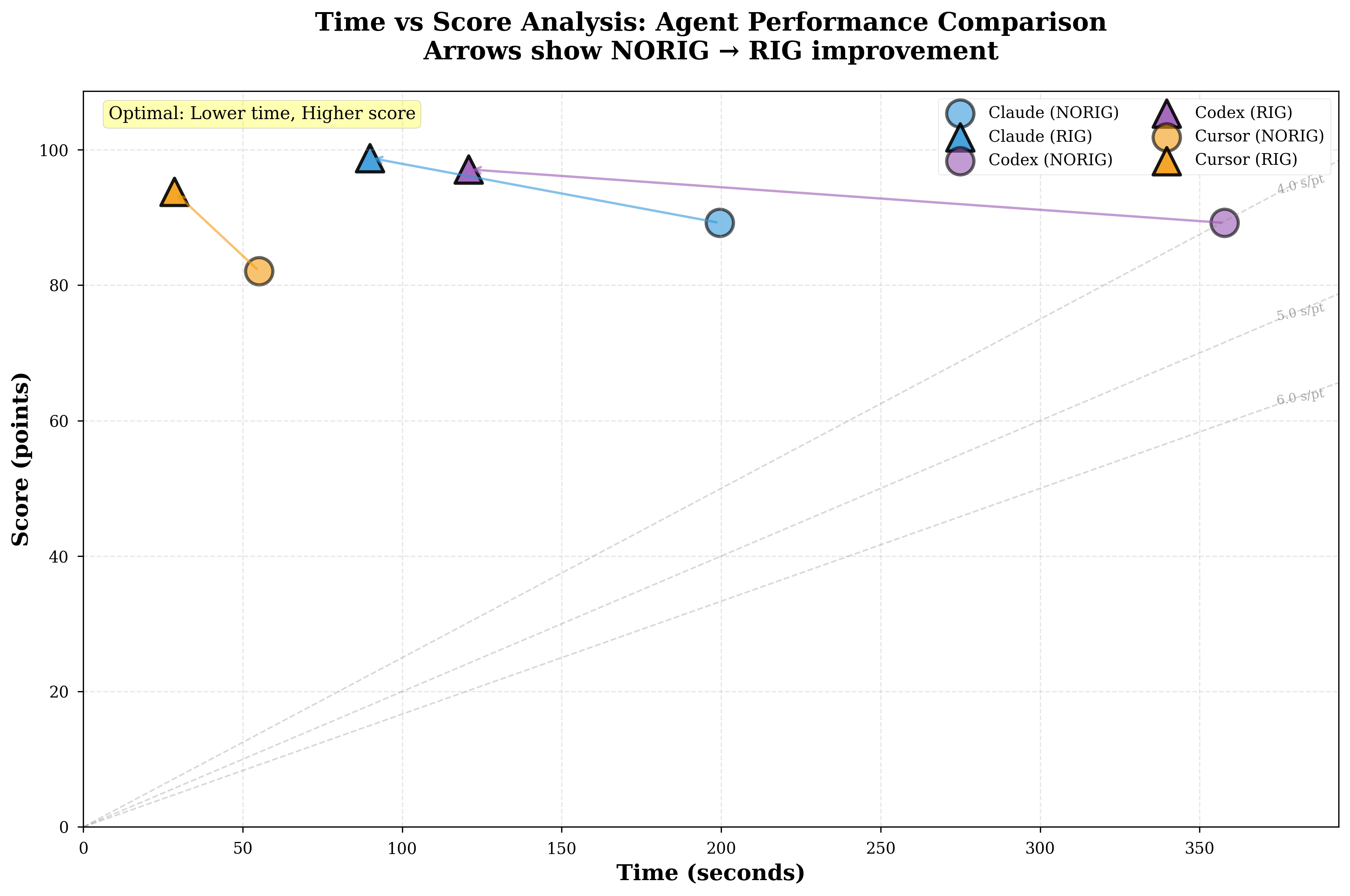}
  \caption{Agent-level mean time versus mean score (averaged across repositories). Circles are NORIG and triangles are RIG, arrows show the NORIG$\rightarrow$RIG shift for each agent. Diagonal lines indicate constant seconds-per-point (efficiency).}
  \label{fig:time-vs-score}
\end{figure}

Figure~\ref{fig:time-vs-score} visualizes the joint effect of RIG on accuracy and runtime at the agent level. For each agent, the plot shows two points: the agent’s mean score and mean completion time averaged across all repositories without RIG (NORIG) and with RIG. Arrows connect each NORIG point to its corresponding RIG point, so upward and leftward arrow movement indicates higher accuracy and shorter completion time, respectively.

The diagonal guide lines correspond to constant seconds-per-point (time divided by score), i.e., constant efficiency. Movement toward a lower-slope line therefore indicates improved efficiency in addition to any score or time change.

Together, Tables~\ref{tab:repo-scores}, \ref{tab:repo-time}, and~\ref{tab:repo-efficiency} provide the global summary of accuracy, latency, and efficiency that the subsequent subsections analyze by complexity, difficulty, category, and agent.

\subsection{Effects by repository complexity}
\label{sec:evaluation-complexity}

Using the normalized complexity scores from Section~\ref{sec:complexity-metric}, we group repositories into low, medium, and high complexity bands: low (score below 30) includes only \texttt{hello\_world}, medium (between 30 and 70 inclusive) includes \texttt{jni}, \texttt{meson}, \texttt{maven}, and \texttt{go}, and high (above 70) includes \texttt{cargo}, \texttt{metaffi}, and \texttt{npm}.

Aggregating the relative accuracy improvements in Table~\ref{tab:repo-scores} by band gives:

\begin{itemize}
  \item Low complexity repositories: average relative accuracy improvement \(1.1\%\).
  \item Medium complexity repositories: average relative accuracy improvement \(11.0\%\).
  \item High complexity repositories: average relative accuracy improvement \(17.4\%\).
\end{itemize}

\begin{figure}[t]
  \centering
  \includegraphics[width=\linewidth]{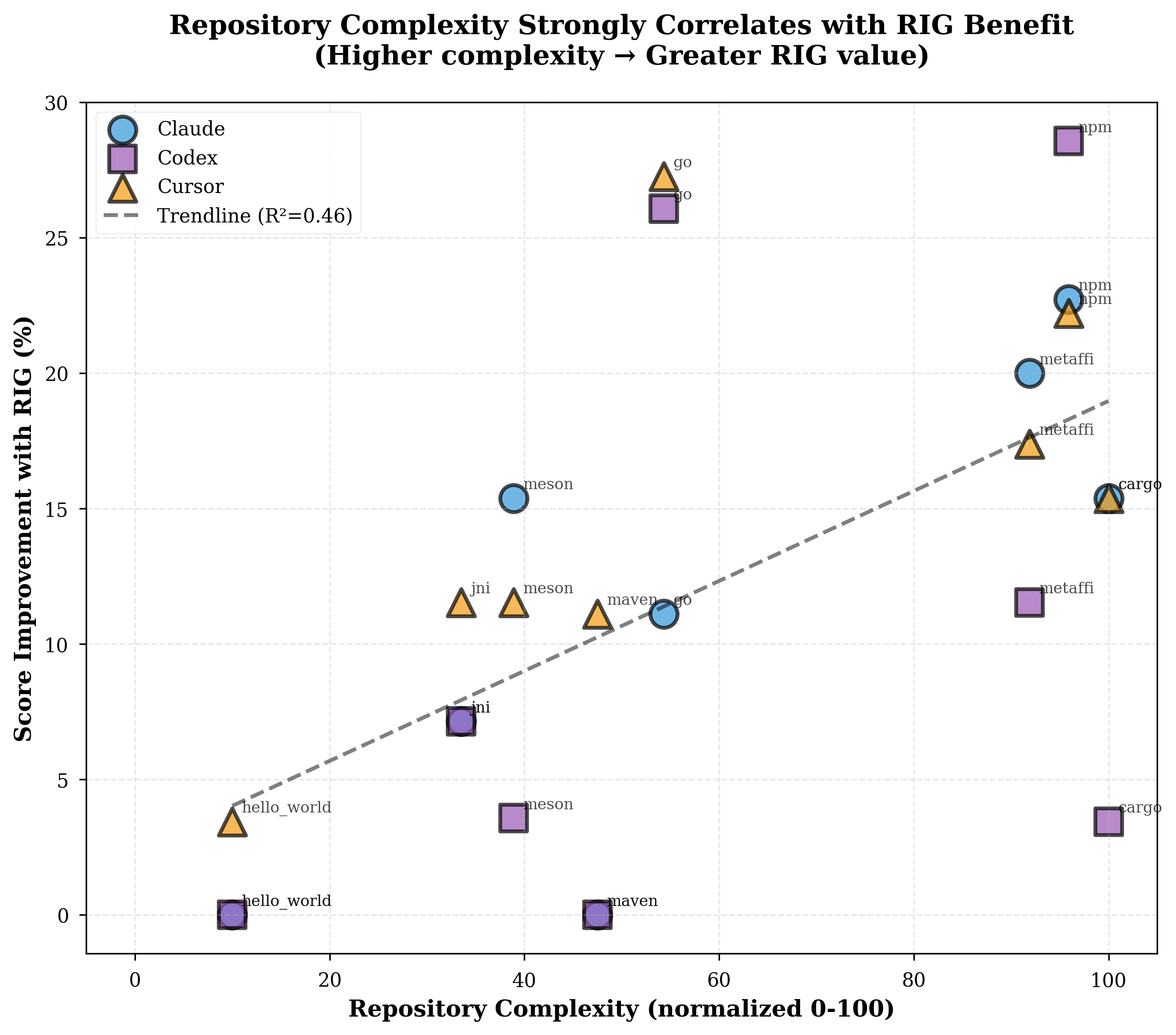}
  \caption{Accuracy improvement with RIG as a function of repository complexity. Each point corresponds to one agent on one repository, and the dashed line is a linear trend fitted across all points.}
  \label{fig:complexity-score-improvement}
\end{figure}

The linear trend line in Figure~\ref{fig:complexity-score-improvement} yields \(R^2 = 0.46\), indicating a moderate positive correlation between repository complexity and score improvement: more complex repositories tend to benefit more from RIG in terms of accuracy.

A similar trend appears for time reduction. Averaging the relative reductions in Table~\ref{tab:repo-time} by band yields 20.8\% for low, 54.6\% for medium, and 64.0\% for high complexity repositories.

\begin{figure}[t]
  \centering
  \includegraphics[width=\linewidth]{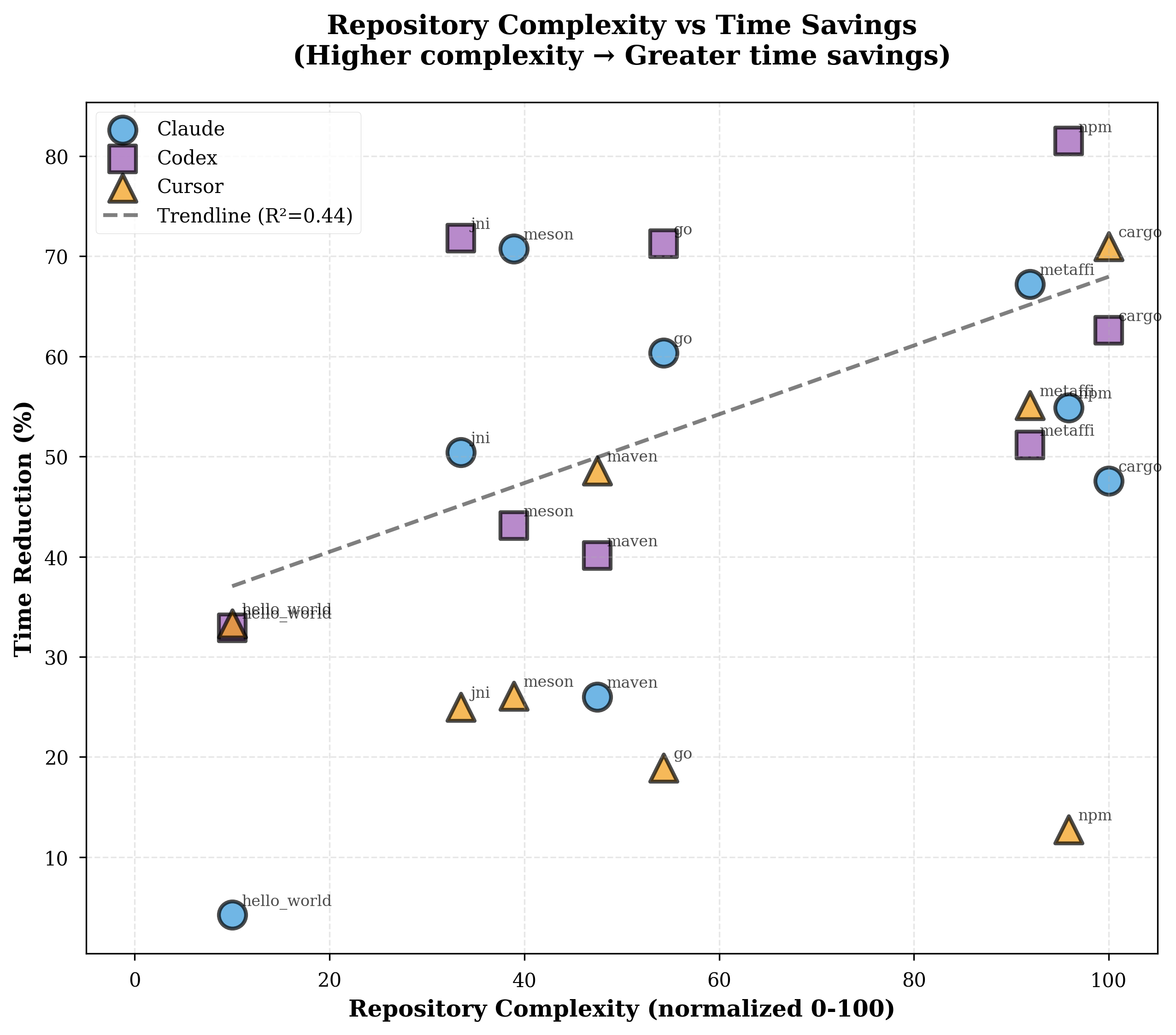}
  \caption{Time reduction with RIG as a function of repository complexity. Each point is one agent-repository pair, the dashed line is a linear trend fitted across all points.}
  \label{fig:complexity-time-reduction}
\end{figure}

The trend line in Figure~\ref{fig:complexity-time-reduction} yields \(R^2 = 0.44\), again indicating a moderate positive correlation between complexity and time savings. Efficiency shows a similar but more pronounced pattern. Aggregating the efficiency improvements in Table~\ref{tab:repo-efficiency} by complexity band gives average reductions in seconds per correct answer of \(21.7\%\) for low complexity repositories, \(58.4\%\) for medium complexity repositories, and \(69.0\%\) for high complexity repositories.

\subsection{Single versus multi-lingual repositories}
\label{sec:evaluation-multilingual}

In addition to complexity, the number of language ecosystems in a repository also affects how much RIG helps. For this analysis we distinguish between repositories that are multi-lingual from the agent's perspective and those that are not, using the \emph{is multi-lingual} column in Table~\ref{tab:repo-efficiency}. We treat pure C/C++ projects as \emph{not} multi-lingual in this analysis and mark the Meson repository explicitly as \texttt{false (C/C++)} to emphasize this point.

Under this classification, \texttt{jni}, \texttt{go}, \texttt{metaffi}, and \texttt{npm} are multi-lingual repositories, in the sense that the agent must reason about multiple distinct language ecosystems and their cross-language wiring. The remaining repositories, including the Meson project, are considered single-language from the agent's point of view. This differs slightly from the complexity metric in Section~\ref{sec:complexity-metric}, where C and C++ are counted as two languages for the purposes of scoring build system complexity. Here we focus on how many distinct ecosystems the agent must mentally juggle.

Using the score improvements from Table~\ref{tab:repo-scores}, the average relative score improvement for the multi-lingual repositories (\texttt{jni}, \texttt{go}, \texttt{metaffi}, \texttt{npm}) is \(17.7\%\), while the average for the single language repositories (\texttt{hello\_world}, \texttt{meson}, \texttt{maven}, \texttt{cargo}) is \(6.6\%\). Thus, multi-lingual repositories see roughly \(2.7\times\) larger accuracy improvements from RIG than single language repositories.

The contrast is also visible for efficiency. Using the efficiency improvements in Table~\ref{tab:repo-efficiency} (reductions in seconds per correct answer), the average efficiency improvement for the multi-lingual repositories is \(70.3\%\), while the average for the single language repositories is \(46.4\%\). When the agent must handle multiple ecosystems, RIG therefore provides not only larger accuracy gains but also substantially larger efficiency gains.

This helps explain the smaller score improvements observed on some medium or high-complexity but single-language repositories such as \texttt{maven} (medium complexity) and \texttt{cargo} (high complexity). In these projects, the build metadata and tooling are already structured around a single language and toolchain, and agents without RIG can often reach relatively high accuracy through conventional exploration. RIG still reduces time and improves efficiency in these cases, but the additional structural guidance has less headroom to translate into very large score gains. In contrast, in multi-lingual repositories with cross-language dependencies, the build and test topology is more intricate, and the explicit cross-language structure encoded in RIG leads to much larger relative improvements.

\subsection{Effects by question difficulty}
\label{sec:evaluation-difficulty}

Questions are grouped into easy, medium, and hard levels as defined in Section~\ref{sec:rig-construction-questions}. Across the eight repositories, there are 240 unique questions in total (30 per repository), with 80 questions per difficulty band. Across the three agents, this corresponds to 720 scored question instances per configuration. The average relative accuracy improvements are:

\begin{itemize}
  \item Easy questions (80 questions): \(3.7\%\) average relative improvement.
  \item Medium questions (80 questions): \(19.0\%\) average relative improvement.
  \item Hard questions (80 questions): \(17.8\%\) average relative improvement.
\end{itemize}

\begin{figure}[t]
  \centering
  \includegraphics[width=0.8\linewidth]{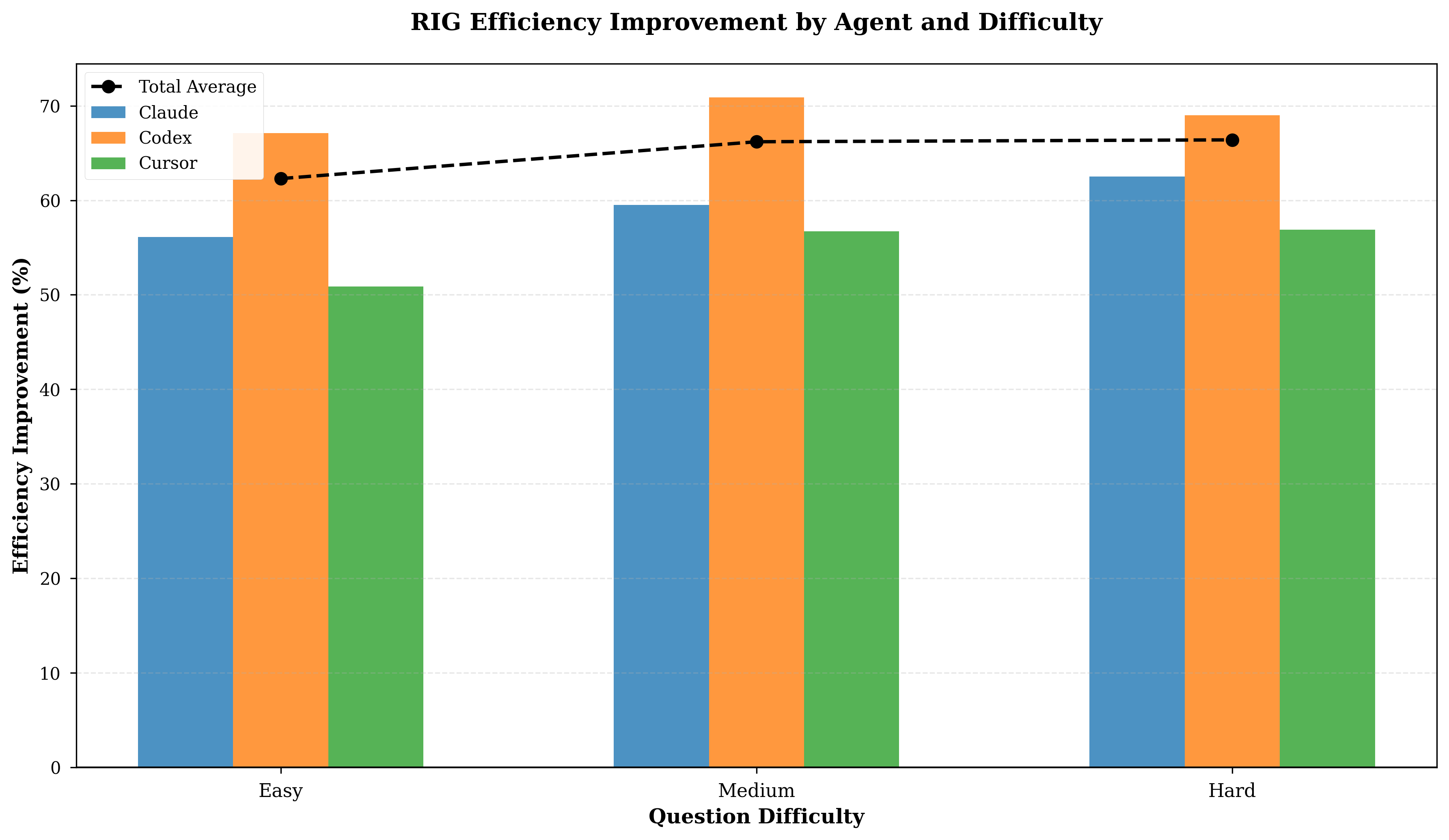}
  \caption{Accuracy improvements with RIG by question difficulty (easy, medium, hard), aggregated across all repositories and agents.}
  \label{fig:difficulty-aggregate}
\end{figure}

In terms of efficiency, measured as seconds per score point, RIG reduces time per score point by \(62.3\%\) on easy questions, \(66.2\%\) on medium questions, and \(66.4\%\) on hard questions, using the efficiency metric defined in Section~\ref{sec:scoring-metrics} and the same per-difficulty aggregates that underlie Figure~\ref{fig:difficulty-aggregate}.

Per-repository difficulty plots (not shown due to space) exhibit the same pattern: easy questions are already near ceiling without RIG and see only small additional gains, while medium and hard questions show substantially larger relative improvements. This matches the design intuition from Section~\ref{sec:rig-construction-questions}: medium questions become straightforward lookups when RIG is available, and hard questions, although they still require non-trivial reasoning, benefit from having the relevant structure made explicit.

\subsection{Agent-specific results}
\label{sec:evaluation-agents}

\begin{figure}[t]
  \centering
  \includegraphics[width=\linewidth]{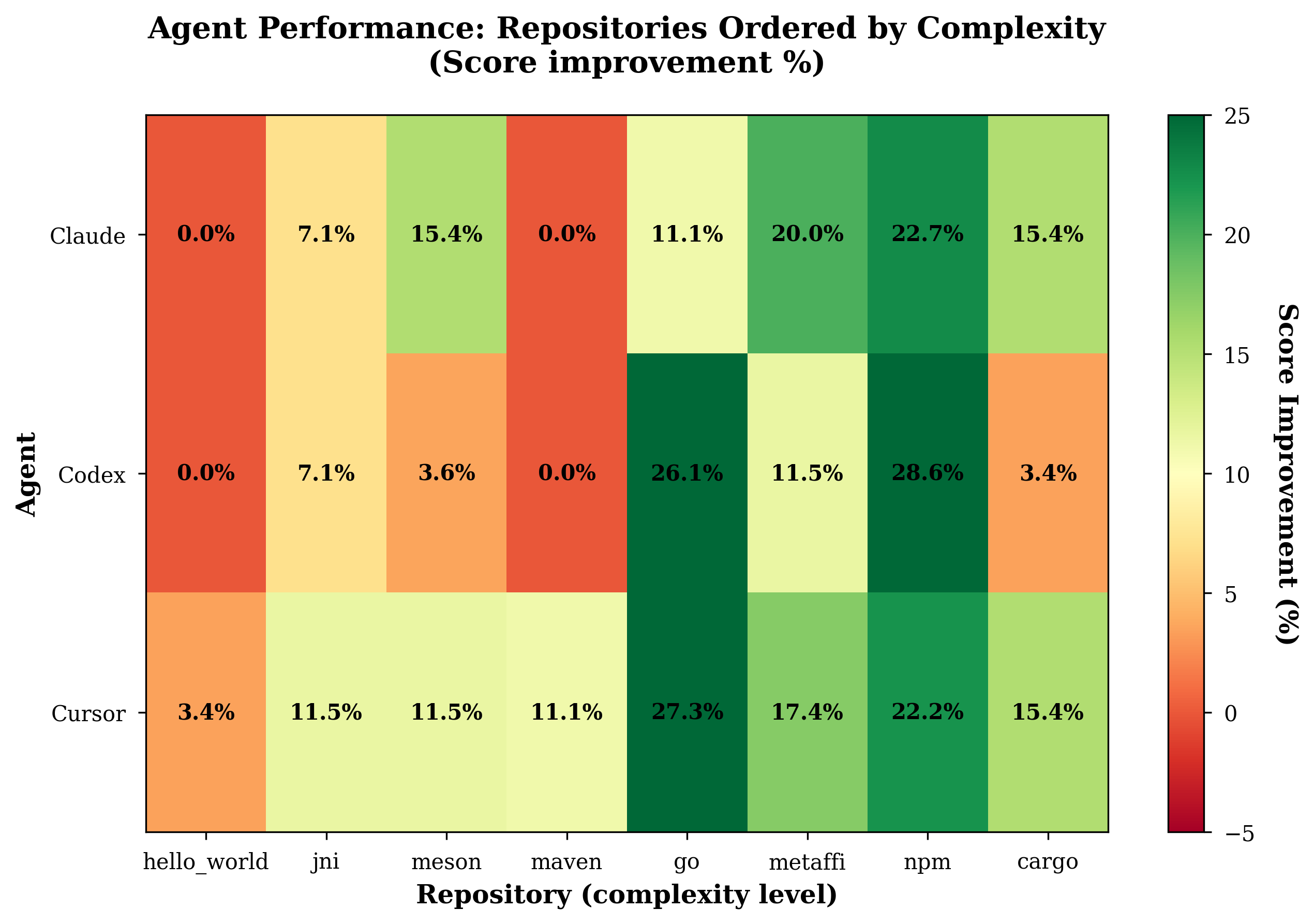}
  \caption{Agent level relative score improvements with RIG across repositories. Cells show relative score improvements (\%) for each agent and repository pair.}
  \label{fig:agent-matrix}
\end{figure}

Figure~\ref{fig:agent-matrix} summarizes relative score improvements for each agent across repositories. Averaging per-repository relative changes yields the following agent-specific effects. We also report each agent's mean accuracy across repositories with and without RIG.

\begin{itemize}
  \item Claude: average \emph{relative} score improvement of \(11.5\%\), average time reduction of \(54.9\%\), mean accuracy without RIG of \(89.2\%\), and mean accuracy with RIG of \(98.8\%\).
  \item Codex: average \emph{relative} score improvement of \(10.0\%\), average time reduction of \(66.2\%\), mean accuracy without RIG of \(89.2\%\), and mean accuracy with RIG of \(97.1\%\).
  \item Cursor: average \emph{relative} score improvement of \(15.0\%\), average time reduction of \(48.3\%\), mean accuracy without RIG of \(82.1\%\), and mean accuracy with RIG of \(93.8\%\).
\end{itemize}

These numbers confirm that each commercial agent benefits from RIG in both accuracy and time, although the exact balance varies by agent. Given the space constraints and the clear pattern in Figure~\ref{fig:agent-matrix}, we use this figure and the aggregate statistics above as the main summary of agent-specific effects and omit additional per-agent plots. Because repositories are ordered by increasing complexity from left to right in Figure~\ref{fig:agent-matrix}, score improvements generally increase toward the right (greener cells), with some variation across agents. For example, \texttt{cargo} shows smaller gains for some agents despite being high complexity, which is consistent with the single versus multi-lingual analysis in Section~\ref{sec:evaluation-multilingual}.

\subsection{Error patterns}
\label{sec:evaluation-errors}

A qualitative inspection of incorrect answers suggests that RIG changes the character of agent errors. Without RIG, many failures are structural. Agents misidentify components, confuse tests with production artifacts, or answer based on incorrect mental models of the dependency graph, often after long sequences of tool calls that still fail to produce a correct global picture.

With RIG, many of these structural failures are reduced. The remaining mistakes more often look like reasoning errors \emph{over the correct graph} (for example incomplete transitive closures or missed reverse dependencies even though the relevant edges are present). We revisit this shift and its implications in the Discussion (Section~\ref{sec:discussion}). A small number of runs exhibit the opposite pattern, where RIG introduces new failure modes or amplifies existing ones. We analyze these ``RIG made it worse'' cases in more detail in Section~\ref{sec:rig-regressions}.

\section{Discussion}
\label{sec:discussion}

The evaluation results in Section~\ref{sec:evaluation-results} support the view of RIG as an architectural map that gives agents a factual backbone about how a repository is built, tested, and assembled, rather than just another block of textual context. In this section we discuss the main patterns that emerge from the experiments, where RIG is most and least helpful, and what this suggests about the role of build and test level structure in repository aware agents.

\subsection{Overall impact of RIG on agent behavior}

Across eight repositories and three commercial agents, providing RIG in the context improves repository level accuracy by an average relative \(12.2\%\) (Table~\ref{tab:repo-scores}) and reduces completion time by \(53.9\%\) on average (Table~\ref{tab:repo-time}). Averaging repository level efficiency improvements in Table~\ref{tab:repo-efficiency} yields a mean efficiency improvement of \(57.8\%\), measured as a reduction in seconds per score point. In other words, agents not only answer more questions correctly when RIG is available, they also complete the same 30-question workload faster and with better efficiency.

The results also show a consistent pattern across agents. Claude, Codex, and Cursor all benefit from RIG in both accuracy and completion time, even though they differ in prompting style, tool integration, and internal architecture (Section~\ref{sec:evaluation-agents}). This suggests that the benefits of RIG are not tied to a specific agent implementation, but arise from the availability of a deterministic build and test level graph that any agent can exploit.

\subsection{Where RIG helps most}

The analysis by question difficulty in Section~\ref{sec:evaluation-difficulty} shows that RIG is most beneficial on medium and hard questions. Easy questions show small gains. They ask for information that is directly visible from a single file or simple directory listing, such as the project name or a file count, and providing RIG yields only a \(3.7\%\) average improvement (already close to 100\% without RIG).

Medium questions, which are designed to be structurally challenging without RIG but directly supported by RIG fields, see the largest gains. These questions include tasks such as identifying the source files of a component or determining whether one component depends on another. Without RIG the agent must parse build system syntax and infer relationships. With RIG it can often answer with a single lookup in the component and dependency collections. The average improvement of \(19.0\%\) on medium questions matches this intended design.

Hard questions still require multi-step reasoning even when RIG is available. They include tasks such as computing build orders or reasoning about reverse dependencies. In these cases RIG removes structural friction by making the dependency graph explicit, but the agent still has to reason over the graph. The average gain of \(17.8\%\) on hard questions shows that RIG helps, but does not trivialize these tasks.

These difficulty trends are consistent with the qualitative error patterns in Section~\ref{sec:evaluation-errors}. RIG reduces friction from repeated parsing and cross-referencing of build and test files, while harder questions can still fail when agents do not correctly reason about component relationships.

\subsection{Multi-lingual repositories and cross-language structure}
\label{sec:multi-lingual-effect}

One of the clearest patterns in the results is the difference between multi-lingual repositories and those that are single language from the agent's point of view. Section~\ref{sec:evaluation-multilingual} shows that multi-lingual repositories (\texttt{jni}, \texttt{go}, \texttt{metaffi}, \texttt{npm}) see an average relative accuracy improvement of \(17.7\%\) and an average efficiency improvement of \(69.5\%\). Single language repositories (\texttt{hello\_world}, \texttt{meson}, \texttt{maven}, \texttt{cargo}) see smaller improvements, with average gains of \(6.6\%\) and \(46.1\%\), respectively.

This aligns with how these repositories differ. Multi-lingual projects force the agent to coordinate several language ecosystems, each with its own build semantics, toolchain, and packaging conventions. The cross-language dependencies that connect these parts are often expressed in terse and idiosyncratic build rules. RIG flattens this complexity into a common graph of components, external packages, runners, and tests, tagged with programming languages but expressed in a single schema.

The Meson repository is an instructive special case. For the complexity metric in Section~\ref{sec:complexity-metric} it counts as multi-lingual, since it combines C and C++ and uses a nontrivial build system. For the multi-lingual analysis in Section~\ref{sec:evaluation-multilingual} it is considered single language from the agent's perspective, because C and C++ share a very similar syntax and toolchain and are often handled by the same mental model and tools. The Meson results show that RIG still improves accuracy and efficiency in such cases, but the effect size is closer to other single language repositories than to the more heterogeneous projects.

The smaller score improvements on high-complexity but single language repositories such as \texttt{maven} and \texttt{cargo} can be interpreted in this light. In these projects the structure that RIG exposes is still useful, but the agent can already reach relatively high accuracy without it because the build metadata is organized around a single language and toolchain. RIG still reduces exploration time and improves efficiency, yet there is less headroom for large score gains. In contrast, in multi-lingual repositories with cross-language dependencies, the build and test topology is more intricate, and the explicit cross-language structure encoded in RIG leads to much larger relative improvements.

\subsection{From structural errors to reasoning errors}

The qualitative error patterns in Section~\ref{sec:evaluation-errors} indicate a shift in the kinds of mistakes agents make when RIG is available. Without RIG, many errors are structural. Agents misidentify which files belong to which component, confuse tests with production artifacts, or make incorrect assumptions about transitive dependencies. These errors often follow long tool traces where the agent visits many directories and build files but fails to synthesize a correct global picture.

With RIG, many of these structural errors are reduced. The graph tells the agent what the buildable components are, which source files they include, which external packages they depend on, and which tests cover which components. The remaining mistakes are more often reasoning errors over this correct structure. For example, an agent may fail to compute a full reverse dependency closure or may overlook an indirect relationship when answering a question about impact or build order.

This shift is aligned with the design goal of RIG. The graph is not meant to replace reasoning. It is meant to remove the need for agents to reverse-engineer the build and test topology from scratch using general-purpose tools. The evaluation suggests that this is indeed what happens in practice.

\subsection{Cases where RIG made things worse}
\label{sec:rig-regressions}

The aggregate results in Section~\ref{sec:evaluation-results} show that RIG never reduces average repository level accuracy: for every repository in Table~\ref{tab:repo-scores}, adding RIG yields a non-negative change in mean score across agents. However, when we examine individual agent-question pairs, there are a small number of cases where a specific agent answers a question correctly without RIG and incorrectly with RIG.

To understand these regressions, we systematically scanned all runs where the with-RIG score for a given agent, repository, and question was lower than the corresponding score without RIG. This yielded four such question-repository pairs. All four involve the same agent (Cursor), and all four are dependency-analysis questions at medium or hard difficulty. Claude and Codex did not exhibit any cases where a previously correct answer became incorrect when RIG was added. For Cursor, these four regressions represent a small but real fraction of its workload: four questions out of 240 per-agent questions in total.

Ideally, we would analyze these cases using direct access to the agent’s internal chain-of-thought. Since this is not supported, we instead combined several indirect methods: replaying the failing questions in isolation, inspecting the detailed tool traces, checking the RIG contents against the ground truth graph, and running small counterfactual experiments (for example, re-asking the same question with slightly different wording or more explicit instructions about transitive dependencies). The patterns that emerge are consistent across the four regressions.

At a high level, the RIG itself is correct in all of these cases: sanity checks against the underlying build systems and manually verified graph queries confirm that the relevant components and edges are present. The problem is how the agent uses the graph. With RIG available, Cursor tends to treat the graph as a high-level index and performs only a shallow traversal (often limited to one or two hops), even when the question explicitly requires a full transitive closure. Without RIG, the same agent sometimes compensates by running more exhaustive tool-based searches (for example, repeated directory walks or build-tool introspection), which, while slower, occasionally lead to a more complete answer.

One representative example comes from the high-complexity npm monorepo. A hard dependency-analysis question asks which services and bundles would be affected if a particular core package failed to build. The RIG for this repository includes a precise dependency graph over components and aggregators: each buildable artifact is a \texttt{Component} node, top-level orchestration scripts are \texttt{Aggregator} nodes, and their \texttt{depends\_on\_ids} edges encode both direct and transitive build relationships. Manual inspection of the graph and automated graph queries confirm that the full set of affected components is reachable by following these edges from the core package.

Without RIG, Cursor answers this question by invoking project-specific tools such as \texttt{npm} or \texttt{pnpm} commands and by scanning multiple \texttt{package.json} files. This exploration is relatively expensive in terms of wall-clock time but eventually reconstructs most of the transitive dependency set, and the answer is scored as correct. With RIG, Cursor instead opens the JSON view, locates the core package component, and then reads only its immediate \texttt{depends\_on\_ids}. The answer it returns includes some directly dependent services but omits components that are affected transitively through aggregators or multi-hop chains. When we re-asked the same question while explicitly instructing the agent to “compute the full transitive closure of dependencies using the RIG graph,” Cursor switched to a breadth-first traversal over \texttt{depends\_on\_ids} and produced the correct set. This indicates that the regression is not due to missing information in RIG, but to a heuristic choice to stop reasoning after a shallow pass over the graph.

A similar pattern appears in a medium-difficulty dependency question in the MetaFFI repository. The question asks which components would be impacted by the failure of a particular core component. The RIG for MetaFFI includes the relevant core component node, all directly and transitively dependent components, and the top-level aggregators that orchestrate them. Without RIG, Cursor spends many tool calls traversing CMake files and build logs and eventually enumerates both direct and indirect dependents, yielding a correct answer. With RIG, it correctly identifies the core component in the JSON and some of its direct dependents, but stops there and does not follow the dependency chain through aggregators into the full set of affected components. Again, manual graph queries over RIG confirm that the missing components are present in the graph and reachable via \texttt{depends\_on\_ids}, the failure is in the agent’s incomplete traversal strategy rather than in the representation.

The remaining regression cases follow the same template. All four occur in dependency-analysis questions that ask about “all affected components” or “the full set of dependents,” and all four involve Cursor with RIG returning a subset of the ground-truth answer while the same agent without RIG either reaches the full set or, in one case, happens to choose a more complete but slower combination of repository tools. In all cases, targeted experiments with more explicit instructions about transitive closure and with simplified prompts confirm that RIG contains the necessary information and that the agent can produce the correct answer when it is nudged to reason more systematically over the graph.

These regressions are rare in absolute terms and confined to a single agent, but they serve as an important reminder that RIG is not a silver bullet. Providing a deterministic, build and test derived graph removes much of the structural uncertainty that agents face when they enter a repository, yet it does not guarantee that they will use that structure optimally. Agents can still adopt shallow heuristics over the graph, stop early, or focus on the wrong subset of nodes. The overall improvements in accuracy and efficiency show that RIG is overwhelmingly helpful, especially on structurally complex and multi-lingual repositories, but the few negative cases highlight that the quality of graph-based reasoning remains a key factor in practice.

\subsection{Implications for repository aware tools and benchmarks}

Taken together, these findings characterize the role that a deterministic build and test level graph can play in repository aware agents.

First, the results support the idea that a build and test centered graph is a useful abstraction boundary. RIG operates above the level of individual files and below the level of informal documentation. It focuses on components and artifacts, their dependencies, and their tests. This abstraction appears to be high enough to generalize across build systems and languages, yet concrete enough to be actionable for agents.

Second, the strong efficiency gains indicate that offloading structural inference to a deterministic extractor is valuable even when accuracy improvements are moderate. In real development workflows, agent latency and the number of interaction steps matter as much as raw correctness. An agent that is much faster at answering the same questions, with higher or equal accuracy, is practically more usable. The results in Sections~\ref{sec:evaluation-repo-time}, \ref{sec:evaluation-complexity}, and~\ref{sec:evaluation-multilingual} show that RIG can offer this kind of efficiency gain, particularly in complex and multi-lingual settings.

Although our benchmark is framed as structured question answering rather than full bug fixing or feature implementation, the questions are intentionally designed to stress the same orientation capabilities that real development workflows rely on. Before an agent can propose a patch or refactor a subsystem, it must first answer questions such as \emph{which components and tests are involved}, \emph{what depends on what}, and \emph{where a change would propagate}. In this sense, structural question answering is a proxy for the ``where should I look?'' phase of development. The improvements we observe suggest that providing RIG can substantially accelerate this orientation step by removing the need to re-derive the build and test topology from scratch, even before any code level reasoning or generation takes place.

Third, the interaction between complexity, multi-lingual structure, and RIG benefit suggests that repository scale evaluations that include high complexity and multi-lingual repositories with realistic build and test setups will better expose the kinds of challenges where architectural knowledge about the build and test graph matters most. Benchmarks that focus only on localized completion tasks in structurally simple projects are unlikely to fully capture these effects.

Finally, the residual error patterns and the small number of true regressions in dependency analysis questions point to concrete opportunities for refinement. The current RIG schema and extractor focus on the build and test topology, not on all possible build system details, and agents are not forced to take full transitive closures when reasoning over the graph. Some build system questions may be outside the intended abstraction layer, or may highlight places where additional schema support or more explicit prompting would be beneficial. Section~\ref{sec:spade-rig} already describes the current schema and extraction logic. The evaluation results in Section~\ref{sec:evaluation-results} and the regression analysis in Section~\ref{sec:rig-regressions} help identify which extensions and refinements are likely to matter most in practice. A detailed discussion of threats to validity appears in Section~\ref{sec:threats-to-validity}.

\section{Threats to Validity}
\label{sec:threats-to-validity}

This section discusses potential threats to the validity of our results and the steps we took, where possible, to mitigate them.

\subsection{Internal validity}

\paragraph{Ground truth RIGs.}
Our conclusions depend on the assumption that the RIGs used in the experiments faithfully represent the build and test structure of each repository. For CMake projects we generate RIG automatically using SPADE, which leverages the CMake File API and related build artifacts (Section~\ref{sec:spade-rig}). During SPADE development we also authored manual ground truth RIGs for the CMake test repositories, including \texttt{metaffi}, using the same schema and Python API. These manual graphs were used iteratively to check that the extractor recovered the same components, dependencies, tests, and external packages. Whenever discrepancies appeared, we either fixed the extractor or corrected the manual RIGs and repeated the comparison. As a result, the CMake RIGs used in the final experiments have been reviewed many times, both through this iterative process and through direct inspection of the corresponding build configurations.

For the non CMake repositories we construct RIGs manually using the same schema and \texttt{RIG} object. While creating these synthetic repositories we maintained separate design documentation that explicitly listed repository specific entities and their relationships, such as components, aggregators, external packages, and dependency edges. We later used this documentation as the specification for the RIGs, and manually reviewed each RIG multiple times to ensure that it matched the documented structure. This documentation is not part of the test repositories themselves and is not available to the agents, so it does not leak answers into the evaluation. In all cases the RIGs were derived from the same build configurations used to compile and test the repositories, cross checked against build outputs, and spot validated by manual inspection. Nevertheless, some residual inaccuracies may remain.

\paragraph{Synthetic repository construction.}
The seven non-\texttt{metaffi} repositories in our corpus are synthetic and were implemented with the assistance of an LLM-based coding assistant. The authors guided the agent, verified and documented each step, and made manual modifications wherever the generated code, build configuration, or tests did not match the intended design. For each synthetic project we maintained separate human-written documentation that specifies its components, aggregators, external packages, and dependency structure, and we used this documentation as the specification when constructing the corresponding RIG. Before including any repository in the evaluation, we manually validated that it builds and runs as intended and that its RIG matches the documented structure. This workflow introduces a potential threat that subtle biases in the AI-assisted construction could shape the repositories in ways that favor or disfavor RIG. However, the RIG graphs are always derived from the final, validated build and test configurations, and all results are reported as within-repository comparisons between runs with and without RIG for the same agent. This reduces the risk that artifacts of the construction process alone drive the observed improvements.

\paragraph{Question design and scoring.}
The thirty evaluation questions per repository (ten easy, ten medium, ten hard) are crafted by the authors. This introduces a risk that the questions inadvertently favor or disfavor the kinds of information RIG encodes. We mitigated this in three ways: first, by using a shared question template across repositories, second, by covering multiple semantic categories (build system, source analysis, testing, dependency analysis, component identification), and third, by defining difficulty levels at design time based on the type and amount of reasoning required, as described in Section~\ref{sec:rig-construction-questions}. Ground truth answers and accepted variants were specified in advance, and all runs were scored using an automated script. We then manually reviewed agent outputs, with particular attention to answers marked as incorrect, to check whether they were actually wrong or simply used an equivalent format or wording. When agents produced correct content in an unexpected but reasonable form, we tightened the question wording, expanded the set of accepted answers, and re-scored all runs for the affected questions. Despite these measures, the benchmark remains a human designed suite rather than a naturally occurring workload.

\paragraph{Manual, interactive agent runs.}
We execute agents in interactive shell mode rather than fully noninteractive batch mode, because noninteractive runs showed unstable behavior and large variance for some agents under repeated trials with identical configurations. Using interactive mode improves stability but introduces a risk of subtle human influence and timing artifacts. To minimize this threat, we automated the evaluation workflow as much as possible. A driver script prepared the exact prompt for each repository and condition and copied it to the clipboard, then launched the agent in interactive mode. The human operator only pasted the prompt once per run and, after the agent produced answers to all thirty questions, terminated the session and recorded the completion time. For Codex and Claude Code we used the agents’ own timing readouts. For Cursor, which does not expose this directly, we measured elapsed time with an external stopwatch started immediately after submitting the prompt and stopped when the run finished. No midrun interventions such as hinting, restarting, or editing prompts were performed. It is also important that we always compare each agent only to itself, with and without RIG, on the same question set. This within agent design means that any systematic timing bias introduced by the interactive procedure (for example overhead from pasting the prompt or starting a stopwatch) affects both conditions in a similar way and is therefore unlikely to distort the reported score improvements, time reductions, or efficiency gains.

\subsection{Construct validity}

\paragraph{Tasks and metrics.}
Our evaluation focuses on structured question answering about repository structure: thirty questions per repository, scored as correct or incorrect and aggregated into accuracy, time, and efficiency (score per second). This is only one slice of what repository-aware agents do in practice. Real-world development tasks often involve code generation, refactoring, bug fixing, or multistep editing sessions, and may use different interaction patterns than our single-prompt configuration. The score and score per second metrics therefore measure the agent’s ability to extract and reason about structural facts efficiently, not overall productivity in a full development workflow.

\paragraph{Abstraction level of RIG.}
RIG is deliberately defined at an architectural level that is grounded in build and test structure. It focuses on components, external packages, runners, and tests, and on the dependency edges between them. It does not include fine-grained code semantics such as control-flow graphs, def use chains, or learned code embeddings. As a result, some kinds of questions that matter to developers, for example those that require detailed reasoning about control flow or data flow inside a component, are outside RIG’s intended scope. Our evaluation design reflects this choice by focusing on questions whose correct answers are either present in RIG or can be derived from it with modest reasoning. The observed gains should therefore be interpreted as gains in structural understanding conditioned on this abstraction boundary.

\paragraph{Efficiency as score per second.}
Because not all commercial agents we use expose token counts, we use wall-clock time and score per second as proxies for efficiency. Wall-clock time depends not only on the agent’s internal computation and tool usage, but also on network latency and local machine load. All runs were carried out on the same machine and network in a compact time window, and evaluations were executed sequentially rather than interleaved with unrelated workloads, in order to reduce environmental variability. For each run we measure end-to-end time from the moment the evaluation prompt is submitted to the agent until all thirty answers are produced, explicitly excluding any agent startup or initialization steps. In other environments with different hardware, network conditions, or rate limiting, absolute times may change even if the relative effects of RIG remain similar. Moreover, score per second does not capture other aspects of usability, such as the number of interaction turns or subjective perceived latency.

\paragraph{Stochasticity and repeated runs.}
LLM-based agents are inherently stochastic, and repeated runs on the same task can yield slightly different outputs and scores. To assess the stability of our findings, we repeated the evaluations for each agent and repository under the same configuration. Individual question scores fluctuated somewhat across runs, as expected, but the aggregate patterns that underpin our conclusions, such as average accuracy gains, time reductions, and efficiency improvements with RIG, remained stable. Since all comparisons are within agent and within repository, and always contrast runs with and without RIG under the same conditions, this residual randomness is unlikely to overturn the main effects reported in Section~\ref{sec:evaluation-results}.

\paragraph{Choice of agents.}
We use three mature commercial agents rather than a custom or minimally engineered baseline. A very weak or immature agent might fail to use tools effectively, so low scores would mainly reflect limitations of the agent itself rather than the absence of RIG. In contrast, the agents in our study are widely used coding assistants with integrated tool support that are already deployed for repository scale work. They are still far from an ideal agent that could always recover the required structure given enough time, but they are strong enough that the benchmark questions are in principle answerable without RIG at a higher cost in time and exploration. Using three different agents also reduces the risk that our findings are specific to a single implementation. The consistent pattern of gains with RIG across all three agents (Section~\ref{sec:evaluation-agents}) suggests that the observed effects are primarily driven by the availability of RIG rather than by idiosyncrasies of one particular agent, although our results do not cover weaker agents or future models with very different tool use behavior.

\subsection{External validity}

\paragraph{Repositories.}
Our corpus consists of eight repositories: seven synthetic but fully buildable projects with controlled complexity, and one real, high-complexity project (MetaFFI). The synthetic repositories are designed to mimic realistic patterns (multi-module builds, microservices, firmware, compilers, and monorepos) while allowing precise control over build complexity and ground truth. This design provides fine-grained control over the independent variable (complexity) but may not cover all patterns found in large industrial codebases. The results are therefore most directly applicable to repositories that resemble our test set in terms of build-system usage, multi-lingual structure, and test organization.

\paragraph{Agents and versions.}
We evaluate three specific commercial agents (Claude, Codex, and Cursor) in concrete tool configurations. The underlying models and toolchains are proprietary and may evolve over time. We record the tool versions and the model names as reported by each agent, but we cannot verify the exact internal model snapshots that were active during the experiments. Future versions of these agents, or other agents with different tool integrations, may exhibit different absolute performance and possibly different sensitivity to RIG. Our results should therefore be interpreted as evidence that RIG can substantially help a range of current commercial agents, rather than as a universal guarantee across all future systems.

\paragraph{Build systems and ecosystems.}
Our RIG extractor is fully automatic only for CMake. RIGs for other build systems (for example Maven, Meson, Cargo, and npm-based setups) are constructed manually using the same schema. This demonstrates that the RIG abstraction can be applied across build systems, but the evaluation does not cover the engineering effort required to build robust extractors for all ecosystems. In addition, the set of languages and frameworks covered by our repositories, while diverse (for example C and C++, Java, Go, Rust, Python, TypeScript, JNI, and WebAssembly), is not exhaustive. Results may differ for repositories that use other build systems or language ecosystems that require different encodings in RIG.

\paragraph{Repositories.}
Our corpus consists of eight repositories: seven synthetic but fully buildable projects with controlled complexity, and one real high-complexity project (\texttt{metaffi}). The synthetic repositories are designed to mimic realistic patterns such as multi-module Maven applications, Go microservices, embedded firmware builds, npm-based monorepos with cross-language bindings, and Cargo-based compilers or interpreters, rather than minimal toy examples. This design provides fine-grained control over the independent variable (build-oriented complexity) while still exposing agents to heterogeneous, multi-component build and test structures.

At the same time, synthetic projects cannot capture the full diversity and idiosyncrasies of large industrial codebases, and they may unintentionally reflect biases introduced by AI-assisted construction. As a result, absolute accuracy levels and raw completion times on our benchmark may differ from what the same agents would achieve on other real-world repositories. However, our evaluation is explicitly within-repository: for each agent and repository we measure the change in accuracy, time, and efficiency when RIG is added to the context, relative to that repository’s own baseline without RIG. The main conclusions therefore concern how RIG affects performance \emph{relative to each repository’s baseline}, not the absolute difficulty of the repositories themselves. This mitigates (but does not eliminate) the threat posed by using synthetic repositories, since the key signal is the with-versus-without-RIG improvement on the same underlying build and test structure. Generalizing the magnitude of these improvements to other repositories, build systems, or organizational settings remains an external validity threat and requires further evaluation.

\subsection{Conclusion validity and reproducibility}

\paragraph{Experimental environment and configurations.}
All experiments were run on the same machine under WSL2 on Windows 11, using Ubuntu~24.04.3 as the guest system. We used Claude Code version 2.0.44 with the ``Sonnet 4.5'' model with thinking enabled, the Cursor agent CLI version 2025.11.06-8fe8a63 with the ``Composer-1'' model, and the OpenAI Codex CLI agent version 0.58.0 with the \texttt{gpt-5.1-codex} model. For each tool we used a fixed prompt template and configuration across all runs. Timing excludes interactive agent startup: the clock starts when the full question prompt is submitted to the already launched agent and stops when the agent has produced answers for all thirty questions. All conditions in the evaluation matrix were executed on this single machine, using the same network and operating system configuration, and runs were carried out sequentially in a narrow time window to reduce environmental drift. We record the tool versions and the model names that the tools report, but we cannot verify the exact internal model snapshots behind these labels.

\paragraph{Non determinism and repetition.}
Commercial LLM-based agents are inherently non deterministic. Sampling, internal tool scheduling, and external factors such as transient network latency can introduce variation, even when prompts and configurations are held fixed. For each combination of repository, agent, and RIG condition we ran at least one full evaluation under controlled settings, and we re-ran selected conditions to check robustness. As expected, individual scores and times varied slightly between repetitions, but the qualitative patterns that matter for this work (for example, that RIG improves accuracy and reduces time, and that improvements grow with complexity and multi-lingual structure) remained stable. We do not have enough repetitions per cell to compute tight confidence intervals, so the reported numbers should be interpreted as indicative summary statistics rather than precise population estimates. The main conclusions rely on consistent directional effects across repositories and agents rather than on fine grained differences between individual percentages.

\paragraph{Release of artifacts.}
To support reproducibility, we release SPADE and the evaluation assets at \url{https://github.com/Greenfuze/Spade}. The public repository includes the SPADE implementation, the RIG schema, ground truth RIG instances for all eight repositories, the synthetic repository sources, the full question sets and accepted answer specifications, and the scoring and orchestration scripts used in the experiments. These assets allow others to reconstruct the evaluation setup, re-run the experiments with the same agents and configurations, or substitute different agents, model versions, or hardware environments to test how the impact of RIG generalizes.

\section{Conclusions and Future Work}
\label{sec:conclusion}

This paper introduced the \emph{Repository Intelligence Graph (RIG)} and SPADE, a deterministic extractor that constructs RIG from build and test artifacts. RIG represents a repository at the level of buildable components, aggregators, runners, external packages, and tests, with explicit dependency and coverage edges and evidence that ties graph facts back to concrete build and test definitions. In our experiments, RIG is serialized into JSON and injected into the agent context before tasks begin, so that agents can treat it as an authoritative structural map instead of reconstructing build and test topology from scratch.

We evaluated three commercial agents on eight repositories that span low, medium, and high build oriented complexity, including a real high complexity project (MetaFFI). Each agent answered thirty structured questions per repository, with and without RIG in its context. Providing RIG improved average score by 12.2\% and reduced average completion time by 53.9\%. Measured as efficiency in seconds per score point, agents spent 57.8\% less time per score point on average when RIG was available, and the per-difficulty reductions fall in a narrow range of 62.3\% to 66.4\% (Sections~\ref{sec:evaluation-repo-time} and~\ref{sec:evaluation-difficulty}). The benefit tends to increase with repository complexity and is larger on average on multi-lingual repositories than on single language ones (Sections~\ref{sec:evaluation-complexity} and~\ref{sec:evaluation-multilingual}). Qualitative error analysis suggests that RIG shifts errors from structural misunderstandings of the repository to reasoning mistakes over a correct graph, which is aligned with its design goal as a structural backbone rather than a reasoning engine. At the same time, the regression analysis in Section~\ref{sec:rig-regressions} shows that RIG is not a silver bullet: in a small fraction of questions it can still nudge agents toward worse answers when they overrely on partial or misinterpreted graph information.

\subsection{Limitations of SPADE and RIG extraction}

The current SPADE prototype demonstrates that it is feasible and useful to derive RIG from existing build and test artifacts, but it also has important limitations.

First, automatic extraction is currently implemented only for CMake-based repositories, and even there it targets the subset of CMake features used in our evaluation corpus. SPADE uses the CMake File API and CTest metadata to construct RIG for the \texttt{hello\_world}, \texttt{jni}, and \texttt{metaffi} projects. For other build systems in this study (Maven, Meson, Cargo, npm-based setups, and Go) we manually author RIGs using the same schema and Python API. This shows that the abstraction itself generalizes across ecosystems, but it does not yet provide an end to end automated pipeline for those systems. Extending SPADE with robust extractors for additional build systems such as Bazel, Gradle, SCons, and Make would require nontrivial engineering effort and careful handling of ecosystem-specific conventions.

Second, even within a supported build system there are cases where the build logic escapes the usual configuration language and the standard metadata SPADE consumes. Many real world projects embed custom scripts, such as shell, Python, or Node.js programs invoked from build rules, or use highly dynamic build descriptions. In such settings, a purely deterministic extractor that relies on the CMake File API and CTest may not see all relevant structure. Our CMake extractor already encounters this in the presence of complex custom commands. When information cannot be recovered reliably from the available artifacts, SPADE records explicit \texttt{UNKNOWN} values or requires manual RIG construction, instead of guessing.

Third, RIG as defined here focuses deliberately on build and test level architecture. It captures components, aggregators, external packages, runners, and tests, and their dependency and coverage edges. It does not include fine grained semantic information about the code, such as control flow or data flow graphs or learned embeddings, and it does not attempt to model higher level requirements or informal documentation. The evaluation shows that this level of abstraction is already useful for structural question answering, but there are likely tasks where combining RIG with richer code level or requirements level representations would be necessary.

\subsection{Towards more generic RIG construction with LLM0}

One of the main practical downsides of SPADE in its current form is that each build system requires explicit extractor support. For CMake this is implemented as a dedicated plugin that understands the CMake File API and CTest outputs. Implementing similarly robust plugins for many build systems and ecosystems would be labor intensive and would still struggle with highly customized or programmatic configurations.

A natural direction for future work is therefore to explore constructing RIG using the pattern matching capabilities of large language models, applied directly to build files, configuration scripts, and related artifacts. The core idea is to let the model infer components, dependencies, tests, and external packages by recognizing recurring patterns in build definitions, rather than relying only on build system specific APIs and schemas.

Directly using a standard LLM in a generative configuration for this purpose raises concerns about randomness and reproducibility. For RIG to serve as a reliable architectural backbone, it must be stable and deterministic enough that repeated extractions on the same repository produce the same graph, and so that errors can be inspected and corrected systematically. To move toward this goal, we consider an approach that minimizes sampling variability when the model is used for pattern extraction. We refer to this as \emph{LLM0}: using an LLM in a configuration that emphasizes predictable pattern matching over creative generation, for example with temperature set to zero, deterministic decoding, and prompts tuned for extraction rather than open ended completion.

Even in the deterministic SPADE pipeline, token cost is not the only concern. Reconstructing RIG from scratch for very large repositories can be non trivial in terms of time and engineering cost, even if no LLM tokens are involved. In practice, a production quality extractor would need incremental update strategies that track which build files and scripts have changed, reuse previous analysis when possible, and validate that the updated graph remains consistent. Designing and evaluating such incremental update mechanisms is an open problem even for purely deterministic extraction.

An LLM0 based extractor magnifies these concerns, because running a large model over an entire repository can be substantially more expensive in both time and tokens. This makes incremental and evidence backed updates even more important. Rather than regenerating the entire graph, an LLM0 pipeline would ideally reanalyze only the changed build and configuration fragments, update the corresponding parts of the RIG, and then check the consistency between the new graph and the previous version.

Evidence becomes even more critical in this setting. Every node and edge inferred by LLM0 should carry evidence that allows humans and downstream tools to verify and, if needed, correct the inference. This supports iterative refinement, where questionable inferences can be traced back to specific build fragments and either confirmed or rejected.

A future LLM0 based extractor could complement or partially replace build system specific plugins in several ways.

\begin{itemize}
  \item As a fallback when no dedicated plugin is available for a given build system, using LLM0 to infer RIG nodes and edges from build files and scripts.
  \item As a cross check for deterministic extractors, comparing their output to an LLM0 derived RIG and flagging discrepancies, with all inferences backed by explicit evidence that points to concrete locations in build or configuration files.
  \item As a tool for handling custom build steps and embedded scripts that lie outside the reach of standard metadata, by interpreting those scripts and mapping their effects into the RIG schema with evidence links.
\end{itemize}

Designing and evaluating such an LLM0 based pipeline raises several open questions. For example, how to represent partial or uncertain inferences in the RIG, how to reconcile conflicting evidence from multiple build artifacts, how to integrate LLM0 reasoning with deterministic extraction so that the resulting graph remains auditable and debuggable, and how to design incremental update policies that balance token cost with freshness. Addressing these questions requires additional experimentation and is left for future work.

\subsection{Broader directions}

Beyond expanding SPADE and exploring LLM0, several broader directions emerge from this work.

First, extending the evaluation to additional tasks would help clarify the scope of RIG’s usefulness. The current benchmarks focus on structured question answering about repository structure. Applying RIG to tasks such as code generation, refactoring, bug fixing, or automated test authoring would test whether the same architectural map improves end-to-end development workflows and not just structural queries. Systematically evaluating RIG on realistic development tasks such as feature implementation, regression fixing, and large-scale refactoring is therefore an important direction for future work.

Second, integrating RIG with other forms of program analysis could enable richer combined views. For example, linking RIG components to static analysis results, control flow or data flow graphs, or learned code embeddings may allow agents to prioritize issues by build and test relevance, or to navigate quickly from a failing test to the most likely root cause.

Third, scaling to larger and more diverse industrial repositories and agent configurations would provide a stronger test of external validity. This includes repositories with hundreds or thousands of components, complex CI and CD pipelines, and heterogeneous tooling, as well as self hosted models or agents with built in graph representations.

Finally, making the evaluation artifacts easily reusable would allow others to rerun, extend, or challenge the results, and to test new agents or extraction methods against the same benchmark. To support this, the public repository at \url{https://github.com/Greenfuze/Spade} includes the RIG schemas, ground truth graphs, question sets, scoring scripts, and supporting documentation, organized so that other researchers can reproduce and adapt the experiments.

In addition, the SQLite backed RIG representation suggests a natural tool interface for agents. Rather than always injecting the full RIG JSON, an agent could query the SQLite store to obtain filtered views that focus on a specific component, test, or dependency slice, and then inject only that smaller JSON into its context.

Overall, the results in this paper suggest that giving agents a deterministic, build and test centered view of a repository is a powerful lever for improving both accuracy and efficiency on repository level understanding tasks, especially in complex and multi-lingual settings. The remaining challenges lie in automating RIG construction across diverse build systems and configurations, and in integrating RIG with complementary representations and analyses to support a wider range of developer tasks.

\section*{Generative AI usage}
We used ChatGPT as a writing assistant for language polishing and for generating Python plotting scripts used to produce figures and supplemental visualizations. All experimental design, data collection, analysis, and conclusions are the authors' work, and we verified the correctness of the generated material. Additional details on our use of LLM based assistance in the study setup and its potential impact are discussed in Section~\ref{sec:threats-to-validity}.

%-----------------------------------------------

\bibliographystyle{IEEEtran}
\bibliography{references}

\end{document}